\documentclass[aps,11pt,pre]{revtex4}
\usepackage[latin1]{inputenc}

\usepackage{epsfig,graphics,subfigure}

\newcommand{\comment}[1]{}

\newcommand{\alphadyn}{\alpha_{\mbox{\scriptsize dyn}}}

\usepackage{amssymb}
\begin{document}
\title{Focused Local Search for Random 3-Satisfiability}

\author{Sakari Seitz}
\affiliation{
  Laboratory for Theoretical Computer Science,
  P.O. Box 5400,
  FI-02015 Helsinki University of Technology, Finland.
  E-mail: {\texttt{firstname.lastname@hut.fi}}.
} 
\affiliation{
  Laboratory of Physics,
  P.O. Box 1100,
  FI-02015 Helsinki University of Technology, Finland.
  E-mail: {\texttt{firstname.lastname@hut.fi}}.
} 
\author{Mikko Alava}
\affiliation{
  Laboratory of Physics,
  P.O. Box 1100,
  FI-02015 Helsinki University of Technology, Finland.
  E-mail: {\texttt{firstname.lastname@hut.fi}}.
} 
\author{
Pekka Orponen}
\affiliation{
  Laboratory for Theoretical Computer Science,
  P.O. Box 5400,
  FI-02015 Helsinki University of Technology, Finland.
  E-mail: {\texttt{firstname.lastname@hut.fi}}.
}

\begin{abstract}
A local search algorithm solving an NP-complete optimisation
problem can be viewed
as a stochastic process moving in an 'energy landscape'
towards eventually finding an optimal solution.
For the random 3-satisfiability problem, the heuristic of
{\em focusing} the local moves on the presently unsatisfied
clauses is known to be very effective:
the time to solution has been observed to grow only linearly
in the number of variables, for a given
clauses-to-variables ratio $\alpha$ sufficiently far below
the critical satisfiability threshold $\alpha_c \approx 4.27$.
We present numerical results on the behaviour of three focused
local search algorithms for this problem,
considering in particular the characteristics
of a focused variant of the simple Metropolis dynamics.
We estimate the optimal value for the ``temperature'' parameter
$\eta$ for this algorithm, such that its linear-time 
regime extends as close to $\alpha_c$ as possible.
Similar parameter optimisation is performed
also for the well-known WalkSAT algorithm
and for the less studied, but very well performing
Focused Record-to-Record Travel method. We observe that
with an appropriate choice of parameters, the linear
time regime for each of these algorithms seems to extend
well into ratios $\alpha > 4.2$ --- much further
than has so far been generally assumed.
We discuss the statistics of solution times for the
algorithms, relate their performance to
the process of ``whitening'',
and present some conjectures on the shape of their
computational phase diagrams.
\end{abstract}
\maketitle
\section{Introduction}

The 3-satisfiability (3-SAT) problem is one of the paradigmatic
NP-complete problems~\cite{GaJo79}. An instance of the problem
is a formula consisting of
$M$ clauses, each of which is a set of three literals,
i.e. Boolean variables or their negations. The goal is to find 
a solution consisting of a satisfying truth assignment
to the $N$ variables, such that each clause contains at least 
one literal evaluating to 'true', provided such an assignment exists. 
In a random 3-SAT instance, the literals comprising each clause
are chosen uniformly at random.

It was observed in~\cite{MiSL92} that random 3-SAT instances
change from being generically satisfiable to being
generically unsatisfiable when the clauses-to-variables
ratio $\alpha = M/N$ exceeds a critical threshold $\alpha_c$.
Current numerical estimates~\cite{BrMZ02} suggest that
this satisfiability threshold is located 
approximately at $\alpha_c \approx 4.267$.
For a general introduction to aspects of the satisfiability
problem see~\cite{DuGP97}.

Two questions are often asked in this context: how to solve
random 3-SAT instances effectively close to the threshold $\alpha_c$,
and what can be said of the point at which various types of algorithms
begin to falter. Recently
progress has been made by the ``survey propagation'' 
method~\cite{BrMZ02, MeZe02, BrZ03}
that essentially iterates a guess about the state of each variable,
in the course of fixing an ever larger proportion of
the variables to their ``best guess'' value.

In this article we investigate the performance of algorithms
that try to find a solution by local moves, flipping
the value of one variable at a time. When the variables to be flipped
are chosen only from the unsatisfied (unsat) clauses, a local search
algorithm is called {\em focusing}. A particularly
well-known example of a focused local search algorithm for 3-SAT
is WalkSAT~\cite{SeKC96}, which makes ``greedy'' and random flips
with a fixed probability. 
Many variants of this and other local search algorithms with different
heuristics have been developed; for a general overview of the
techniques see~\cite{AaLe97}.

Our main emphasis will be on a focused variant of the
standard Metropolis dynamics~\cite{Metr53} of spin systems,
which is also the base for the well-known simulated
annealing optimization method. When applied to the 3-SAT
problem, we call this dynamics the Focused Metropolis Search
(FMS) algorithm. We also consider the Focused Record-to-Record
Travel (FRRT) algorithm~\cite{Duec93,SeOr03}, which imposes
a history-dependent upper limit for the maximum energy allowed
during its execution.

The space of solutions to 3-SAT instances 
slightly below $\alpha_c$ is known
to develop structure. Physics-based methods from mean-field -like
spin glasses imply through replica methods that the
solutions become ``clustered'', with a threshold of 
$\alpha \approx 3.92$~\cite{MeZe02}. 
The clustering implies that if measured e.g.\ by Hamming distance
the different solutions in the same cluster are close
to each other. A possible consequence of this is the existence
of cluster ``backbones'', which means that in a cluster
a certain fraction of the variables is fixed, while the
others can be varied subject to some constraints. The
stability of the replica ansatz becomes crucial for higher $\alpha$,
and it has been suggested that this might become important at 
$\alpha \approx 4.15$ \cite{MoPR04}. The energy landscape in which
the local algorithms move is however a finite-temperature
version. The question is how close to $\alpha_c$ one can get
by moving around by local moves (spin flips) and focusing
on the unsat clauses. 

Our results concern the optimality of this strategy for 
various algorithms and $\alpha$. First, we demonstrate that
WalkSAT works in the ``critical'' region up to $\alpha > 4.2$
with the optimal noise parameter $p \approx 0.57$.
Then, we concentrate on the numerical performance of
the FMS method. In this case,
the solution time is found to be linear in $N$
within a parameter window $\eta_{min} \leq \eta \leq \eta_{max}$,
where $\eta$ is essentially the Metropolis temperature parameter.
More precisely, we observe that within this window, the median and all
other quantiles of the solution times normalised by $N$ seem to
approach a constant value as $N$ increases. A stronger condition
would be that the distributions of the solution times be concentrated
in the sense that the ratio of the standard deviation and the average
solution time decreases with increasing $N$. While numerical studies of
the distributions do not indicate heavy tails that would
contradict this condition,  we can of course not at present prove
that this is rigorously true.
The existence of the $\eta$ window implies that for too 
large $\eta$ the algorithm becomes
entropic, while for too small $\eta$ it is too greedy,
leading to freezing of degrees of freedom
(variables) in that case.

The {\em optimal} $\eta = \eta_{opt}(\alpha)$,
i.e.\ that $\eta$ for which the solution time median is lowest,
seems to increase with increasing $\alpha$.
We have tried to extrapolate this behaviour towards
$\alpha_c$ and consider it possible that the algorithm
works, in the median sense, linearly in $N$ all the way up
to $\alpha_c$. This is in contradiction with some earlier
conjectures. All in all we
postulate a phase diagram for the FMS algorithm based on
two phase boundaries following from too little or too
much greed. Of this picture, one particular phase space
point has been considered before~\cite{BaHW03,SeMo03}, since it 
happens to be the case that for $\eta=1$ the FMS coincides
the so called Random Walk method~\cite{Papa91},
which is known to suffer from metastability at 
$\alpha \approx 2.7$ \cite{BaHW03, SeMo03}.

The structure of this paper is as follows: In Section 2 we present as
a comparison the WalkSAT algorithm, list some of the known results
concerning the random 3-SAT problem, and test the WalkSAT for varying
$p$ and $\alpha$ values. In Section 3 we go through the FMS algorithm
in detail, and present extensive numerical simulations with it.
Section 4 presents as a further comparison similar data for
the FRRT algorithm. In Section 5 we discuss how the
notion of whitening~\cite{Pari02} is related to the behaviour of FMS
with low $\eta$. Finally, section 6 finishes the paper with a summary.

\section{Local Search for Satisfiability}

It is natural to view the satisfiability problem as
a combinatorial optimisation task, where the goal for
a given formula $F$ is to minimise the objective function
$E = E_F(s) = $ the number of unsatisfied clauses
in formula $F$ under truth assignment $s$.
The use of local search heuristics in this context was
promoted e.g.\ by Selman et al.\ in~\cite{SeLM92} and
by Gu in~\cite{Gu92}. Viewed as a spin glass model, $E$
can be taken to be the energy of the system.

\comment{
\begin{figure}\center
\begin{verbatim}
GSAT(F):
  s = random initial truth assignment;
  while flips < max_flips do
    if s satisfies F then output s & halt, else:
    - find a variable x whose flipping causes
      largest decrease in E (if no decrease is
      possible, then smallest increase);
    - flip x.
\end{verbatim}
\caption{The GSAT algorithm.}
\label{fig:gsat}
\end{figure}

\begin{figure}\center
\begin{verbatim}
NoisyGSAT(F,p):
  s = random initial truth assignment;
  while flips < max_flips do
    if s satisfies F then output s & halt, else:
    - with probability p, pick a variable x
      uniformly at random and flip it;
    - with probability (1-p), do basic GSAT move:
      - find a variable x whose flipping causes
        largest decrease in E (if no decrease is
        possible, then smallest increase);
      - flip x.
\end{verbatim}
\caption{The NoisyGSAT algorithm.}
\label{fig:noisygsat}
\end{figure}
}

Selman et al.\ introduced in~\cite{SeLM92} the simple greedy
GSAT algorithm, whereby at each step the variable to be flipped
is determined by which flip yields the largest decrease in $E$,
or if no decrease is possible, then smallest increase.
This method was improved in~\cite{SeKC96} by augmenting the
simple greedy steps with an adjustable fraction $p$
of pure random walk moves, leading to the algorithm
NoisyGSAT.

\begin{figure}\center
\begin{verbatim}
WalkSAT(F,p):
  s = random initial truth assignment;
  while flips < max_flips do
    if s satisfies F then output s & halt, else:
    - pick a random unsatisfied clause C in F;
    - if some variables in C can be flipped without
      breaking any presently satisfied clauses,
      then pick one such variable x at random; else:
    - with probability p, pick a variable x
      in C at random;
    - with probability (1-p), pick some x in C
      that breaks a minimal number of presently
      satisfied clauses;
    - flip x.
\end{verbatim}
\caption{The WalkSAT algorithm.}
\label{fig:wsat}
\end{figure}

In a different line of work,
Papadimitriou~\cite{Papa91} introduced the
very important idea of {\em focusing} the search to
consider at each step only those variables that appear
in the presently {\em unsatisfied} clauses.
Applying this modification to the NoisyGSAT
method leads to the WalkSAT~\cite{SeKC96} algorithm
(Figure~\ref{fig:wsat}), which is the currently
most popular local search method for satisfiability.

In~\cite{SeKC96}, Selman et al.\ presented some 
comparisons among their new NoisyGSAT and WalkSAT algorithms,
together with some other methods. These experiments were
based on a somewhat unsystematic set of benchmark formulas
with $N \leq 2000$ at $\alpha \approx \alpha_c$, but nevertheless
illustrated the significance of the focusing idea, since in
the results reported, WalkSAT outperformed NoisyGSAT by several
orders of magnitude.

More recently, Barthel et al.\ \cite{BaHW03} performed systematic
numerical experiments with Papadimitriou's original Random Walk
method at $N = 50,000$, $\alpha = 2.0\dots 4.0$.
They also gave an analytical explanation for a transition in the
dynamics of this algorithm at $\alphadyn \approx 2.7$, already
observed by  Parkes~\cite{Park02}. When $\alpha < \alphadyn$,
satisfying assignments are generically found in time that is linear
in the number of variables,
whereas when $\alpha > \alphadyn$ exponential time is required.
(Similar results were obtained by Semerjian and Monasson
in~\cite{SeMo03}, though with smaller experiments ($N = 500$).)
The reason for this dynamical threshold phenomenon seems
to be that at $\alpha > \alphadyn$ the search equilibrates
at a nonzero energy level, and can only escape to
a ground state through a large enough random fluctuation.
A rate-equation analysis of the method~\cite{BaHW03} 
yields a very well matching approximation of
$\alphadyn \approx 2.71$.\footnote{Our numerical experiments
with the Random Walk algorithm suggest that
its dynamical threshold is actually somewhat lower,
$\alphadyn \approx 2.67$.}
See also~\cite{SeMo04} for further analyses of the
Random Walk method on random 3-SAT.

WalkSAT is more powerful than the simple Random Walk,
because in it focusing is accompanied by other heuristics.
However, it is known that the behaviour of WalkSAT is quite
sensitive to the choice of the noise parameter $p$. E.g.\
Parkes \cite{Park01} experimented with the algorithm using
a noise value $p = 0.3$ and concluded that with this setting
the algorithm works in linear time at least
up to $\alpha = 3.8$. Even this result is not the best
possible, since it has been estimated~\cite{Hoos02,HoSt99, WeSe02}
that for random 3-SAT close to the satisfiability transition
the optimal noise setting for WalkSAT is $p \approx 0.55$.
(Actually our numerical experiments, reported below,
suggest that the ideal value is closer to $p \approx 0.57$.)

These positive results notwithstanding, it has been
generally conjectured (e.g.\ in~\cite{BrMZ02})
that no local search algorithm can work in linear time
beyond the clustering transition at $\alpha_d \approx 3.92$.
In a series of recent experiments, however,
Aurell et al.~\cite{AuGK04} concluded that with a proper
choice of parameters, the median solution time of WalkSAT 
remains linear
in $N$ up to at least $\alpha = 4.15$, the onset of 1-RSB symmetry
breaking. Our experiments, reported below, indicate that
the median time in fact remains linear even beyond that,
in line with our previous results~\cite{SeOr03}.

\begin{figure}\center
\subfigure[Complete data]{
   \epsfig{file=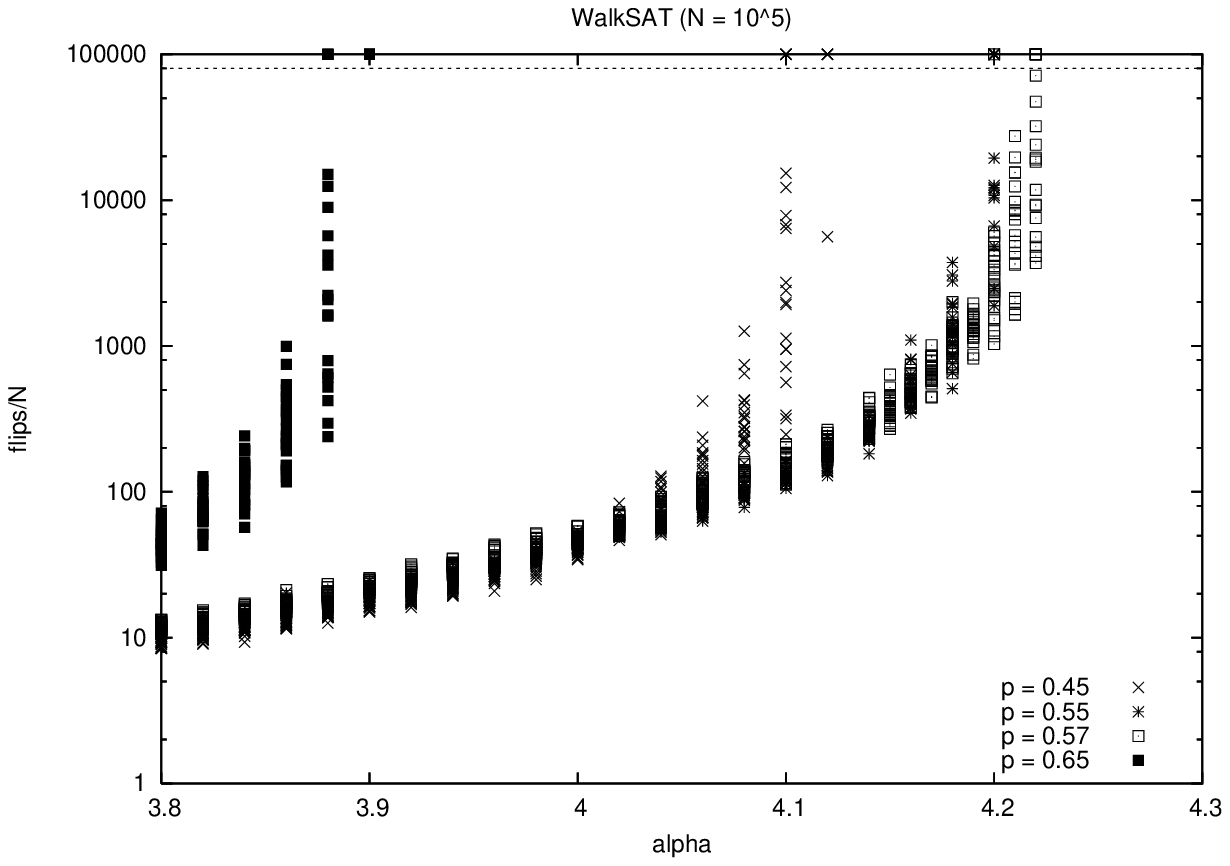,width=0.45\linewidth}
}
\subfigure[Medians and quartiles]{
   \epsfig{file=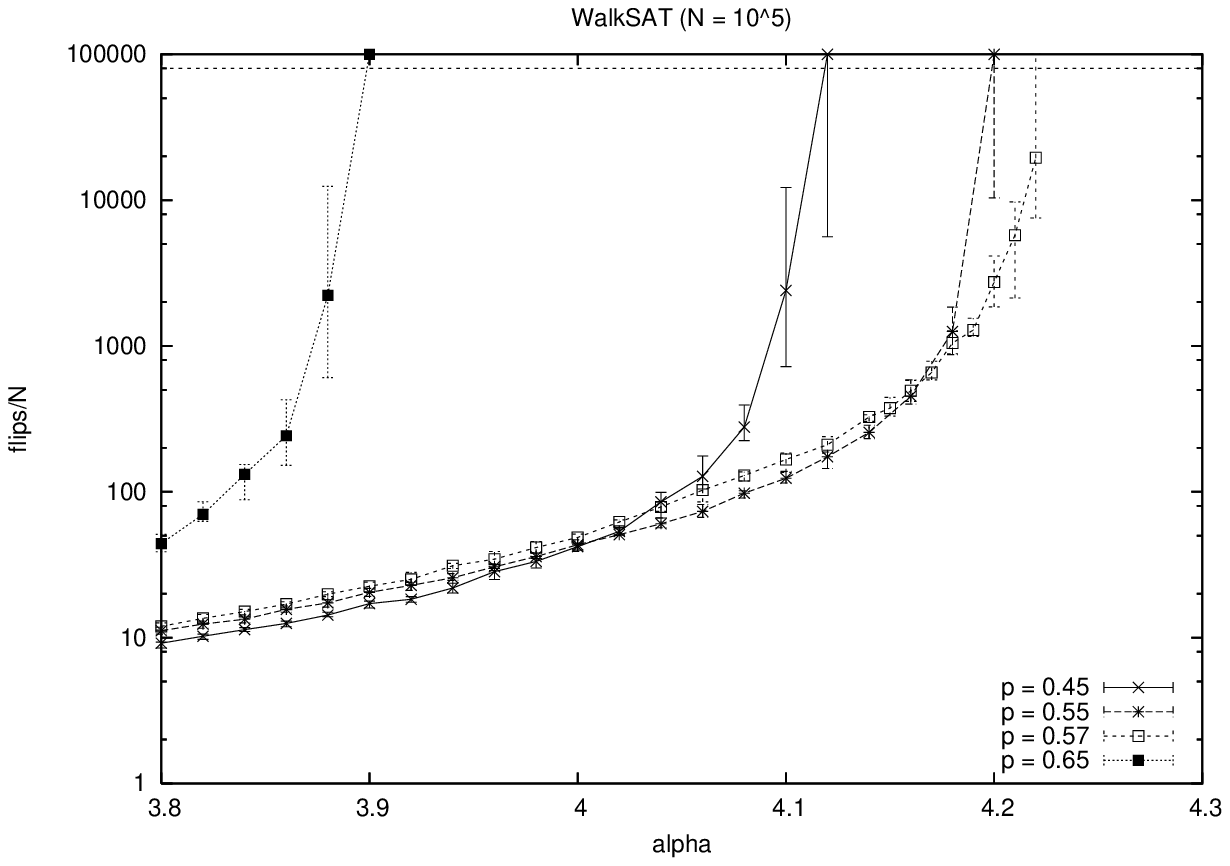,width=0.45\linewidth}
}
\caption{Normalised solution times for WalkSAT, $\alpha = 3.8\ldots 4.3$.}
\label{fig:wsat_scan}
\end{figure}

Figure~\ref{fig:wsat_scan} illustrates our experiments
with the WalkSAT algorithm,~\footnote{Version 43, downloaded
from the Walksat Home Page at
{\tt http://www.cs.washington.edu/homes/kautz/walksat/},
with its default heuristics.}
on randomly generated formulas of size $N = 10^5$,
various values of the noise parameter $p$,
and values of $\alpha$ starting from $3.8$ and
increasing at increments of $0.2$ up to $4.22$. For each
$(p,\alpha)$-combination, 21 formulas were generated, and for
each of these the algorithm was run until either a satisfying
solution was found or a time limit of $80000\times N$ flips
was exceeded. Figure~\ref{fig:wsat_scan}(a) shows the solution times
$t_{sol}$,
measured in number of flips normalised by $N$, for each generated
formula. Figure~\ref{fig:wsat_scan}(b) gives the medians and quartiles
of the data for each value of $\alpha$.

As can be seen from the figures, for value $p = 0.45$ of
the noise parameter WalkSAT finds satisfying
assignments in roughly time linear in $N$,
with the coefficient of linearity increasing
gradually with increasing $\alpha$, up to approximately
$\alpha = 4.1$ beyond which the distribution of solution
times for the algorithm diverges. For $p = 0.55$, this linear
regime extends further, up to at least $\alpha = 4.18$, but
for $p = 0.65$ it seems to end already before $\alpha = 3.9$. For the
best value of $p$ we have been able to experimentally determine,
$p = 0.57$, the linear regime seems to extend up to even
beyond $\alpha = 4.2$.

\begin{figure}\center
\subfigure[$\alpha = 4.15$]{
   \epsfig{file=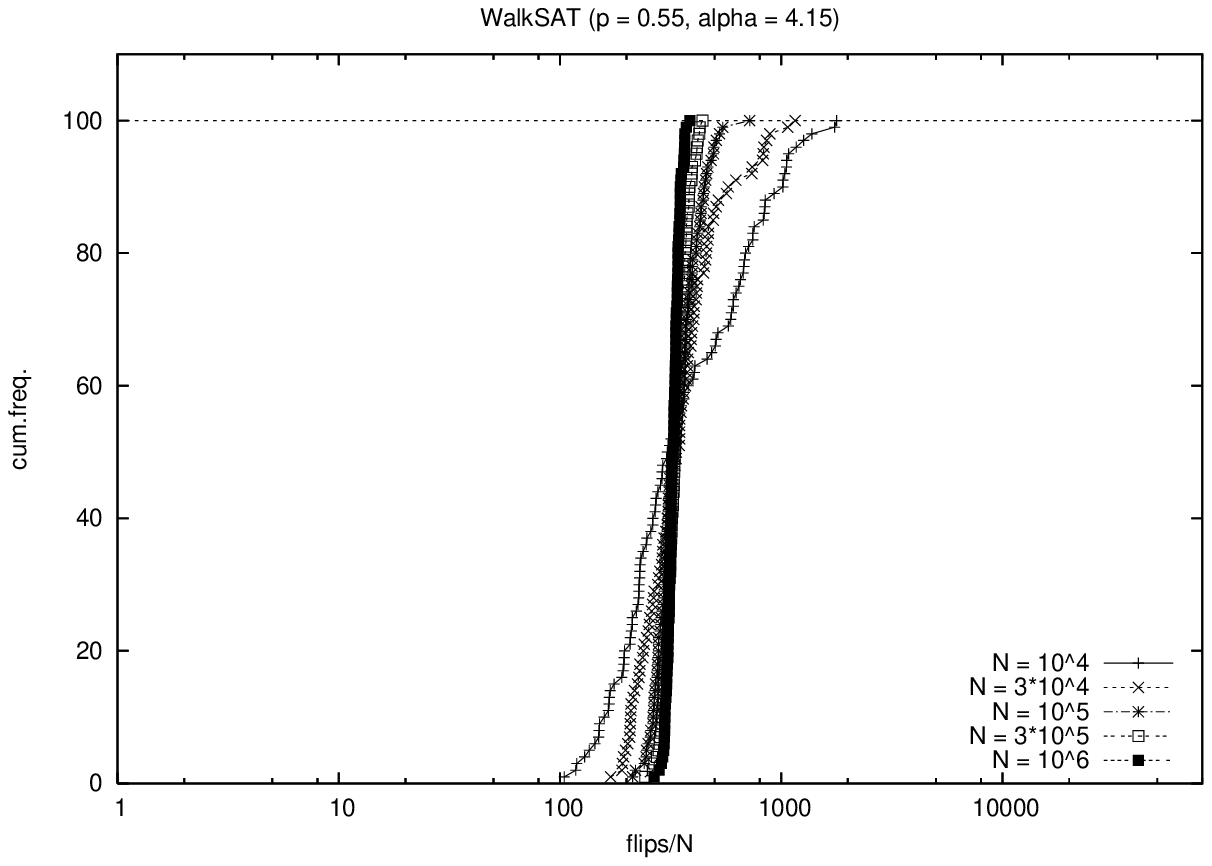,width=0.45\linewidth}
}
\subfigure[$\alpha = 4.20$]{
   \epsfig{file=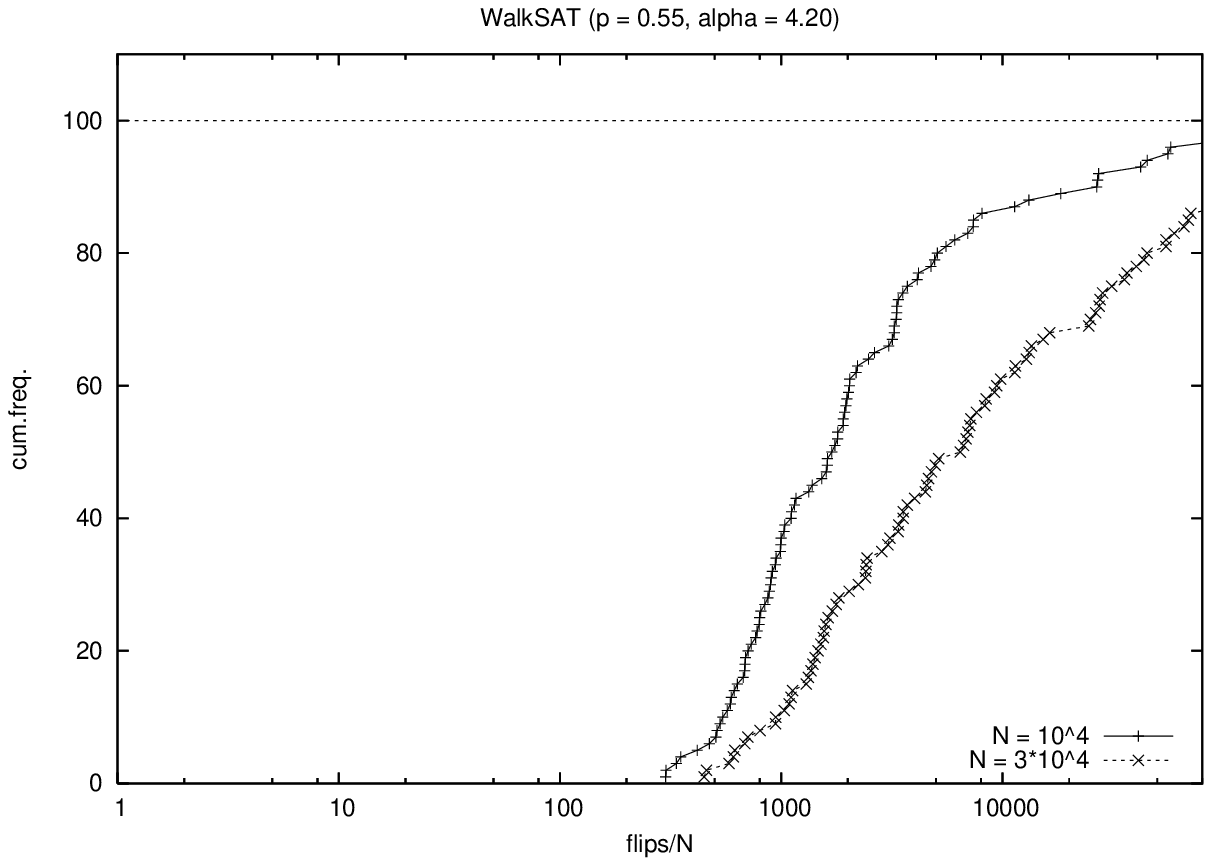,width=0.45\linewidth}
}
\caption{Cumulative solution time distributions for
WalkSAT with $p = 0.55$.}
\label{fig:wsat55_dist}
\end{figure}

To investigate the convergence of the solution time distributions,
we tested the WalkSAT algorithm with $p = 0.55$ at $\alpha = 4.15$
and $\alpha = 4.20$, 
in both cases with randomly generated sets of 100 formulas
of sizes $N = 10^4, 3\times 10^4, 10^5, 3\times 10^5$ and $10^6$.
Figure~\ref{fig:wsat55_dist} shows the cumulative 
distributions of the solution times normalised by $N$
achieved in these tests.
As can be seen, for $\alpha = 4.15$ the distributions are
well-defined, with normalised medians and all other quantiles
converging to a finite value for increasing $N$.
However, for $\alpha = 4.20$, the distributions seem
to diverge, with median values increasing with increasing $N$.\footnote{
For $\alpha = 4.20$, the tests for $N = 10^5$ and larger were not
completed, because the solution times consistently overran the
time limit of $80000 \times N$ flips, and consequently the
test runs were exceedingly long yet uninformative.}

\begin{figure}\center
   \epsfig{file=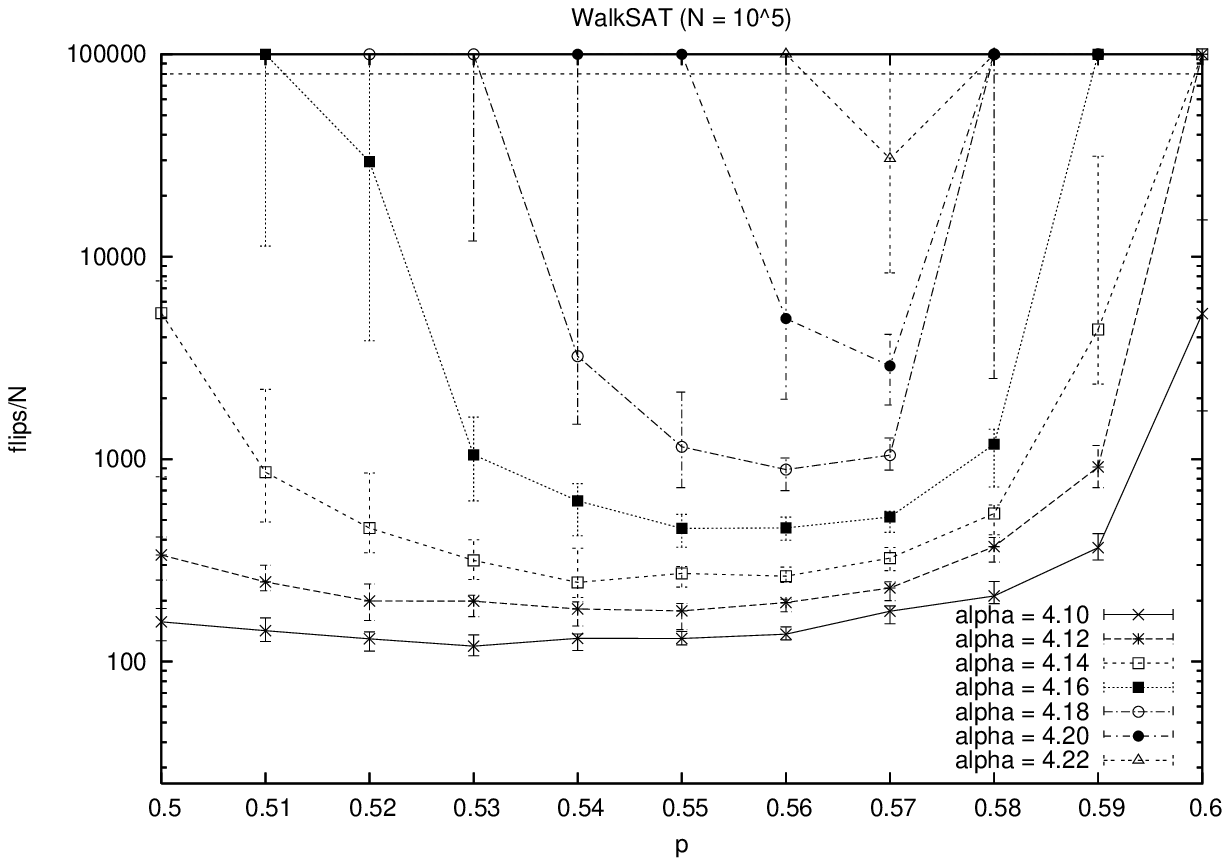,width=0.45\linewidth}
\caption{Normalised solution times for WalkSAT with $\alpha = 4.10 \dots 4.22$,
$p = 0.50\dots 0.60$.}
\label{fig:wsatmin_med}
\end{figure}

\begin{figure}\center
   \epsfig{file=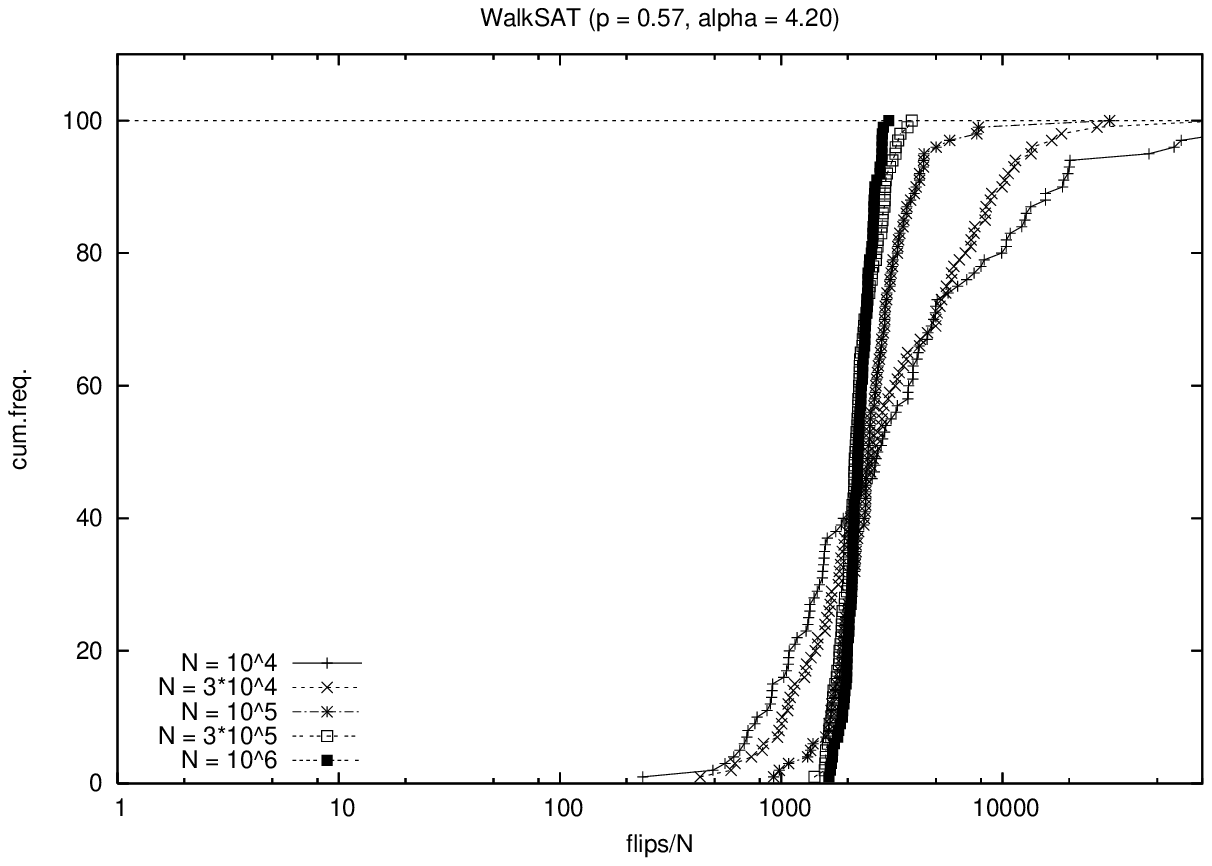,width=0.45\linewidth}
\caption{Cumulative solution time distributions for
WalkSAT with $p = 0.57$, $\alpha = 4.20$.}
\label{fig:wsat57_dist}
\end{figure}

In order to estimate the optimal value of the WalkSAT noise
parameter, i.e.\ that value of $p$ for which the linear time regime
extends to the biggest value of $\alpha$, we generated test
sets consisting of 21 formulas, each of size $N = 10^5$, for
$\alpha$ values ranging from $4.10$ to $4.22$ at increments of
$0.02$, and for each $\alpha$ for $p$ values ranging from $0.50$
to $0.60$ at increments of $0.01$. WalkSAT was then run on
each resulting $(\alpha, p)$ test set; the medians and
quartiles of the observed solution time
distributions are shown in Figure~\ref{fig:wsatmin_med}. 
The data suggest that the optimal value of the noise
parameter is approximately $p = 0.57$.
Figure~\ref{fig:wsat57_dist} shows the empirical solution
time distributions at $\alpha = 4.20$ for $p = 0.57$.
In contrast to Figure~\ref{fig:wsat55_dist}(b),
now the quantiles seem again to converge to a finite value for
increasing $N$, albeit more slowly than in the case of the simpler
$\alpha = 4.15$ formulas presented in Figure~\ref{fig:wsat55_dist}(a).

\section{Focused Metropolis Search}

From an analytical point of view, the WalkSAT algorithm
is rather complicated, with its interleaved greedy and
randomised moves. 
Thus, it is of interest to investigate the
behaviour of the simpler algorithm obtained by
imposing the focusing heuristic
on a basic Metropolis dynamics~\cite{Metr53}.

\begin{figure}\center
\begin{verbatim}
FMS(F,eta):
  s = random initial truth assignment;
  while flips < max_flips do
    if s satisfies F then output s & halt, else:
      pick a random unsatisfied clause C in F;
      pick a variable x in C at random;
      let x' = flip(x), s' = s[x'/x];
      if E(s') <= E(s) then flip x, else: 
         flip x with prob. eta^(E(s')-E(s)).
\end{verbatim}
\caption{The Focused Metropolis Search algorithm.}
\label{fig:fms}
\end{figure}

We call the resulting algorithm, outlined in
Figure~\ref{fig:fms}, the {\em Focused Metropolis Search}
(FMS) method. The algorithm is parameterised by a number
$\eta$, $0 \leq \eta \leq 1$, which determines the probability
of accepting a candidate variable flip that would lead to
a unit increase in the objective function $E$. 
(Thus in customary Metropolis dynamics terms, $\eta = e^{1/T}$,
where $T$ is the chosen computational temperature.
Note, however, that detailed balance does not hold with 
focusing.)

\begin{figure}\center
\subfigure[Complete data]{
   \epsfig{file=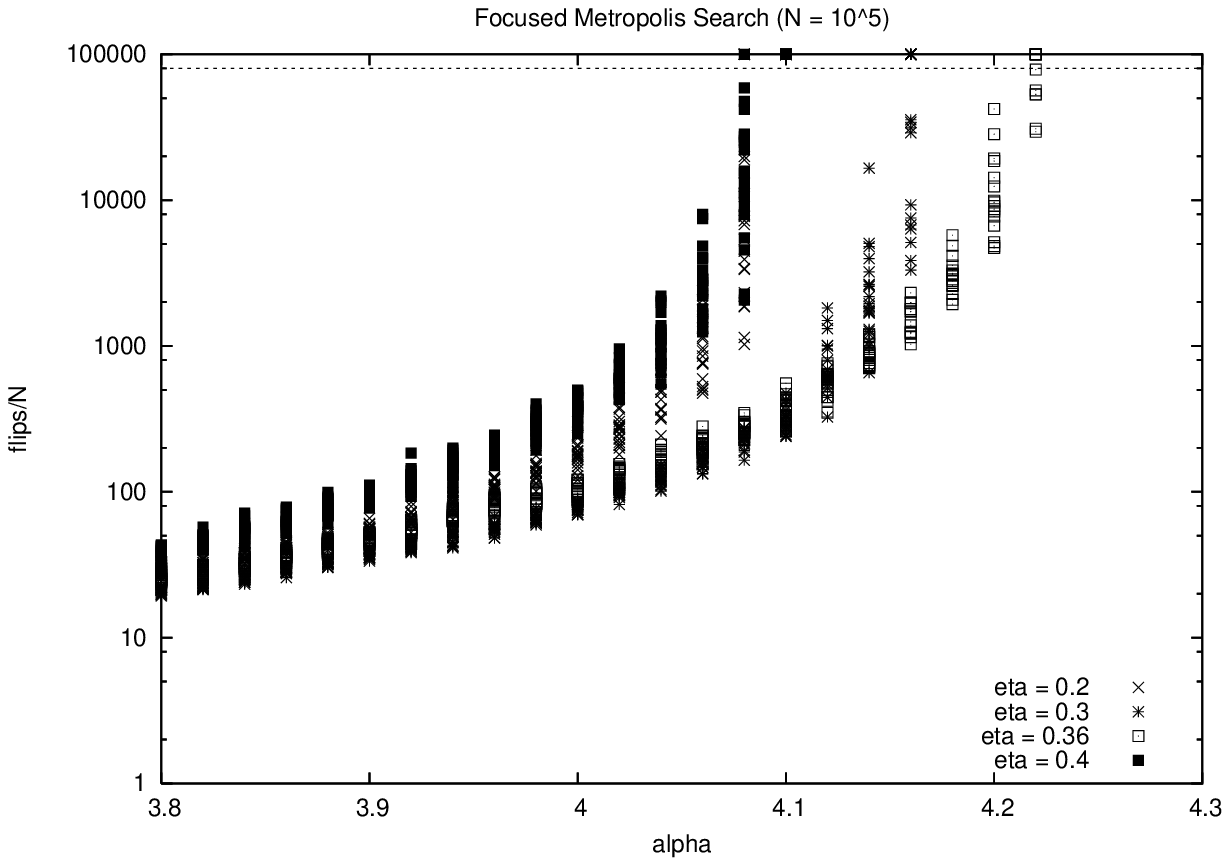,width=0.45\linewidth}
}
\subfigure[Medians and quartiles]{
   \epsfig{file=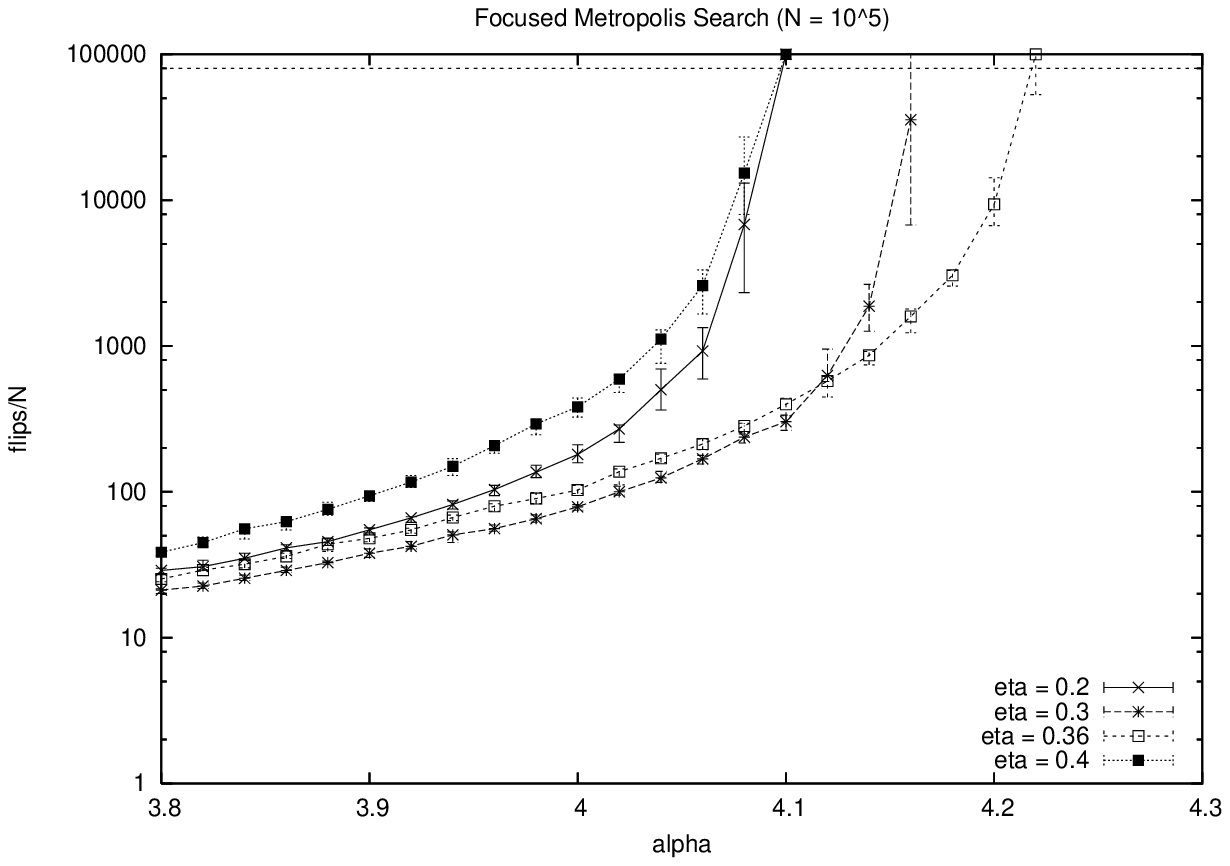,width=0.45\linewidth}
}
\caption{Normalised solution times for FMS, $\alpha = 3.8\ldots 4.3$.}
\label{fig:fms_scan}
\end{figure}

We repeated the data collection procedure of
Figure~\ref{fig:wsat_scan} for the FMS algorithm with
various parameter values. The results for $\eta = 0.2, 0.3, 0.4$
and $\eta = 0.36$ (the best value we were able to find)
are shown in Figure~\ref{fig:fms_scan}; note that
also rejected flips are here included in the flip counts.
As can be seen,
the behaviour of the algorithm is qualitatively quite
similar to WalkSAT.
For parameter value $\eta = 0.2$, the linear time regime seems
to extend up to roughly $\alpha = 4.06$, for $\eta = 0.3$ up to
at least $\alpha = 4.14$, and for $\eta = 0.36$ even beyond
$\alpha = 4.20$; however for $\eta = 0.4$ again only
up to maybe $\alpha = 4.08$.
\begin{figure}\center
\subfigure[$\alpha = 4.00$]{
   \epsfig{file=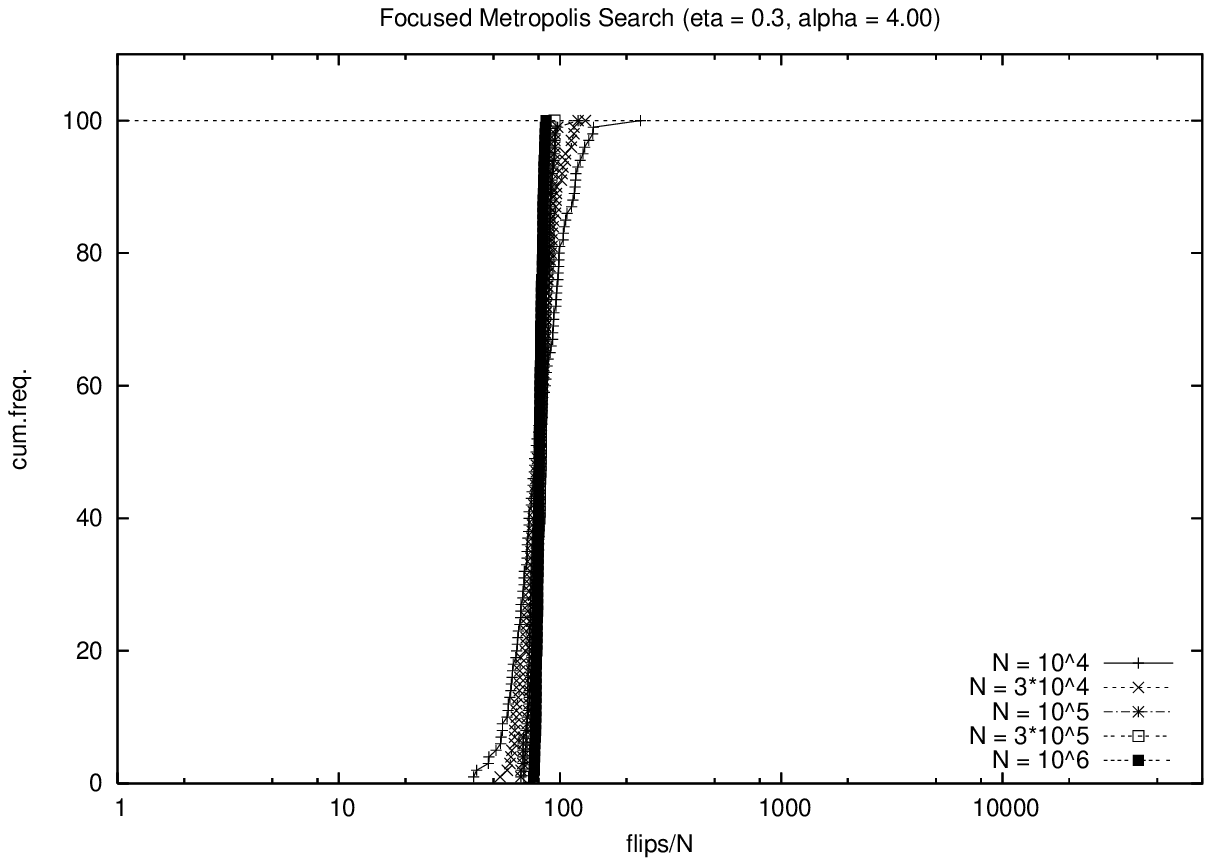,width=0.45\linewidth}
}
\subfigure[$\alpha = 4.10$]{
   \epsfig{file=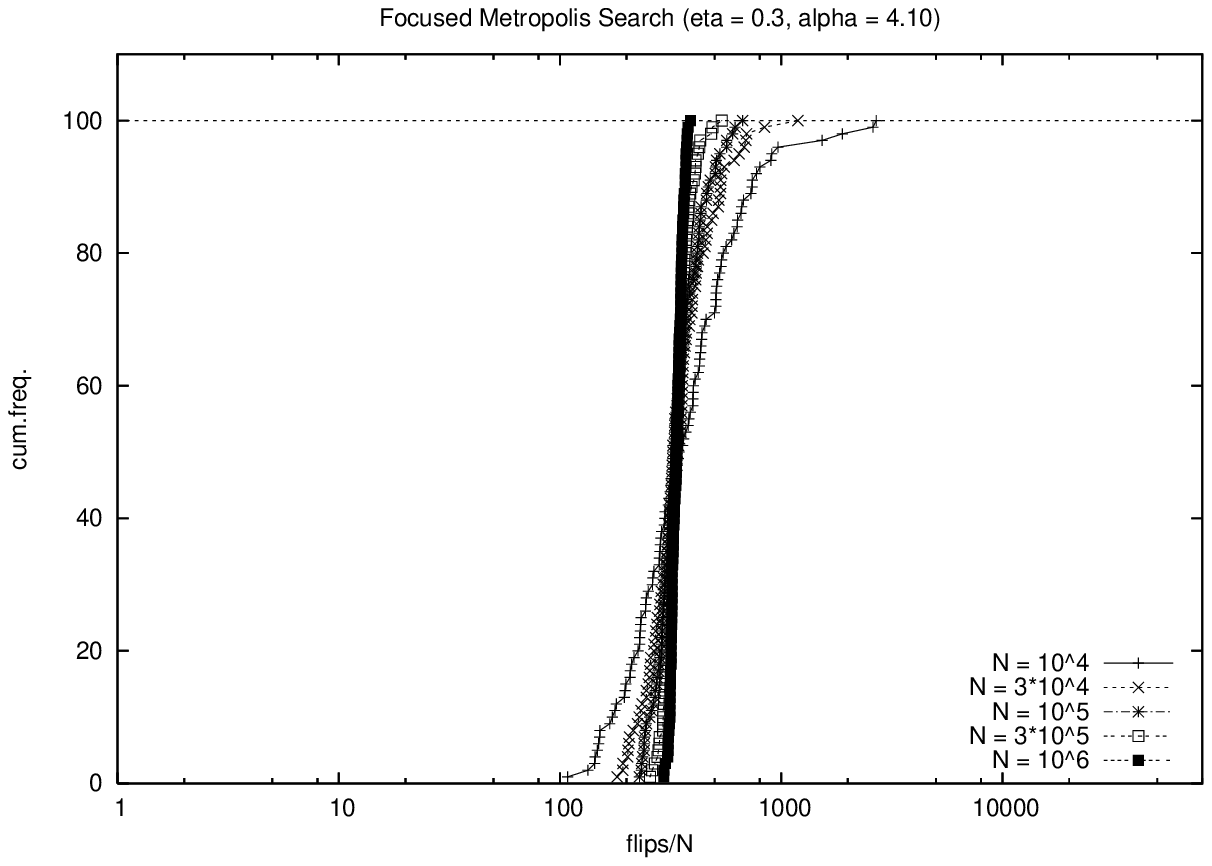,width=0.45\linewidth}
}
\caption{Cumulative solution time distributions for
FMS with $\eta = 0.3$.}
\label{fig:fms03_dist}
\end{figure}
To test the convergence of distributions, we determined
the empirical cumulative distributions of FMS solution times
for $\eta = 0.3$ at $\alpha = 4.0$ and $\alpha = 4.1$,
in a similar manner as in Figure~\ref{fig:wsat55_dist}.
The results are shown in Figure~\ref{fig:fms03_dist}.

\begin{figure}\center
   \epsfig{file=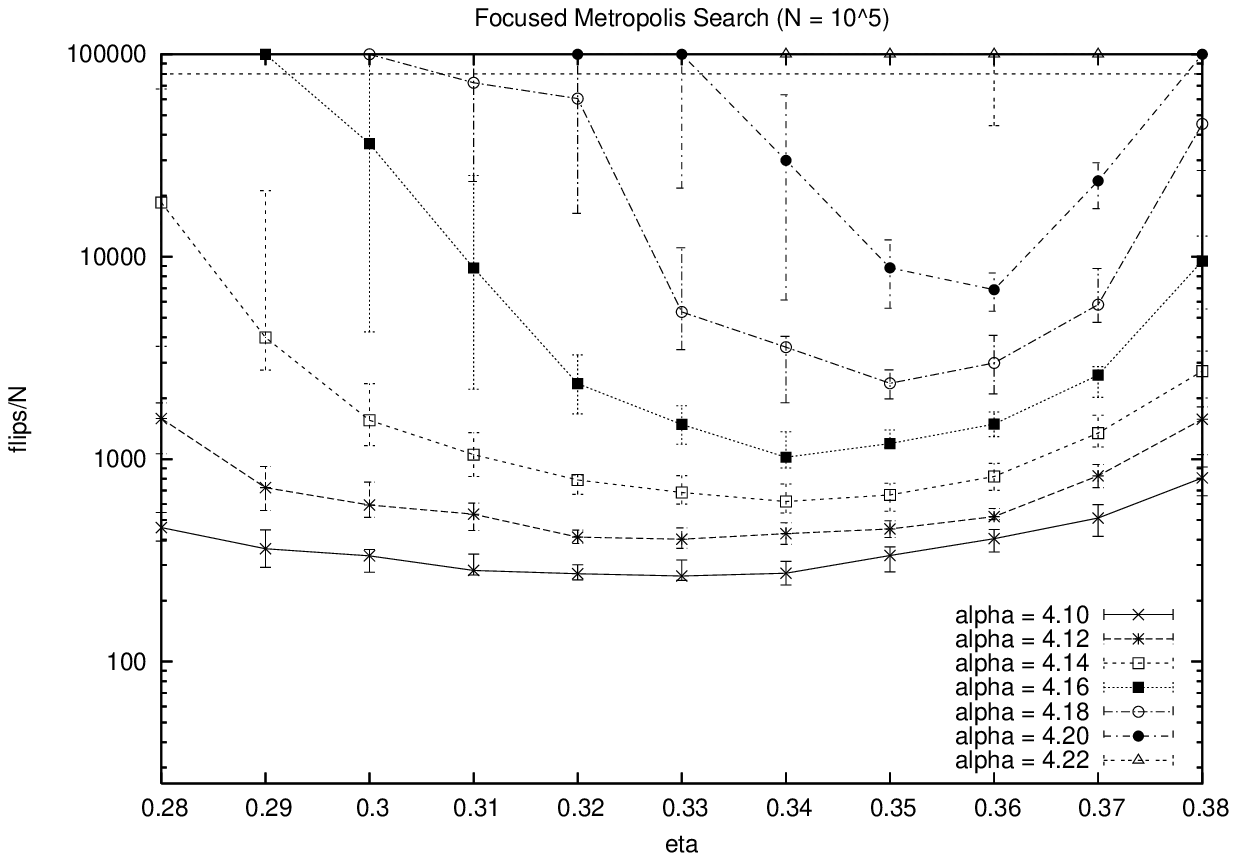,width=0.45\linewidth}
\caption{Normalised solution times for FMS with $\eta = 0.28\dots 0.38$,
$\alpha = 4.10 \dots 4.22$.}
\label{fig:fmsmin_med}
\end{figure}

To determine the optimal value of the $\eta$ parameter
we proceeded as in Figure~\ref{fig:wsatmin_med}, mapping
out systematically the solution time distributions of the
FMS algorithm for $\alpha$ increasing from $4.10$ to
$4.22$ and $\eta$ ranging from $0.28$ to $0.38$. The
results, shown in Figure~\ref{fig:fmsmin_med}, suggest
that the optimal value of the parameter
is approximately $\eta = 0.36$.

\begin{figure}\center
   \epsfig{file=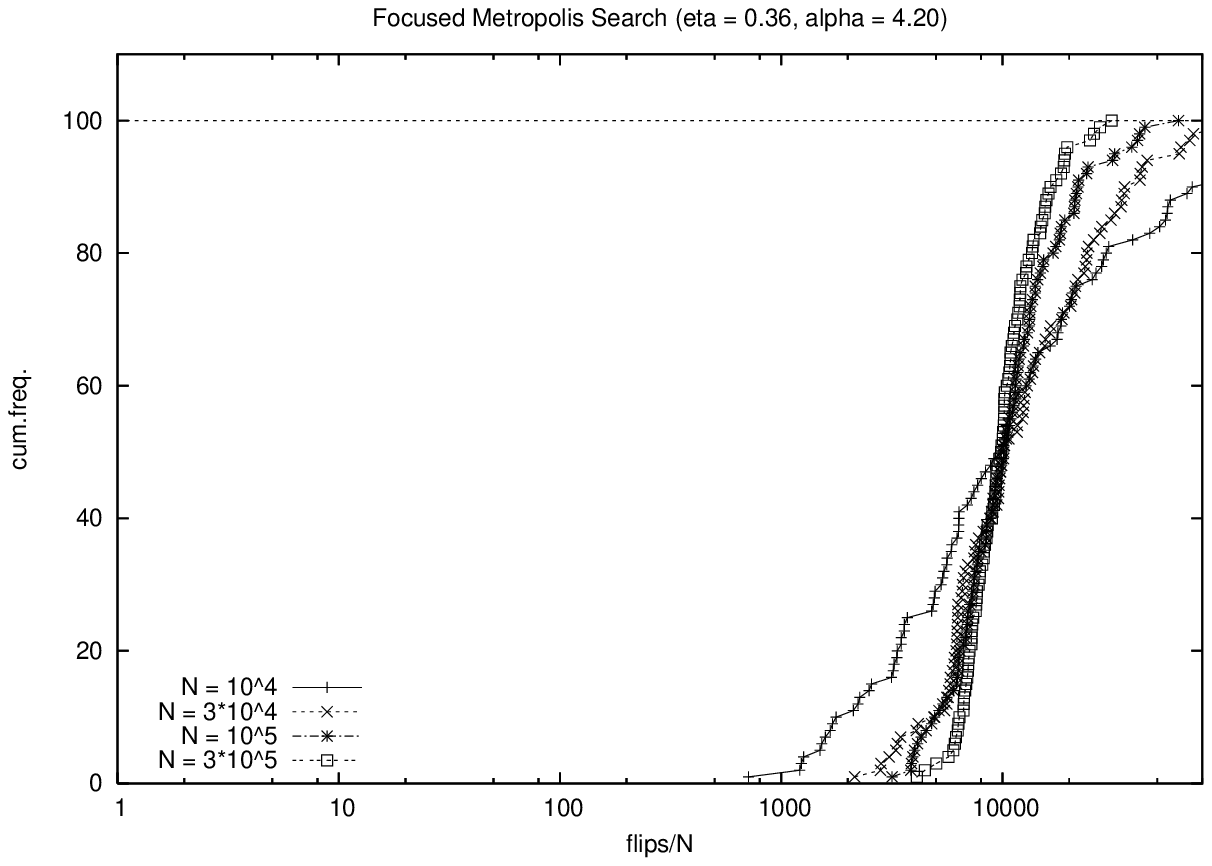,width=0.45\linewidth}
\caption{Cumulative solution time distributions for
FMS with $\eta = 0.36$, $\alpha = 4.20$.}
\label{fig:fms036_420}
\end{figure}

In order investigate the algorithm's behaviour
at this extreme of its parameter range, we determined the
empirical cumulative distributions of the FMS solution 
times for $\eta = 0.36$ at $\alpha = 4.20$.
The results, shown in Figure~\ref{fig:fms036_420},
suggest that even for this high value of $\alpha$,
the FMS solution times are linear
in $N$, with all quantiles converging to 
a finite value as $N$ increases.

\begin{figure}\center
\subfigure{
   \epsfig{file=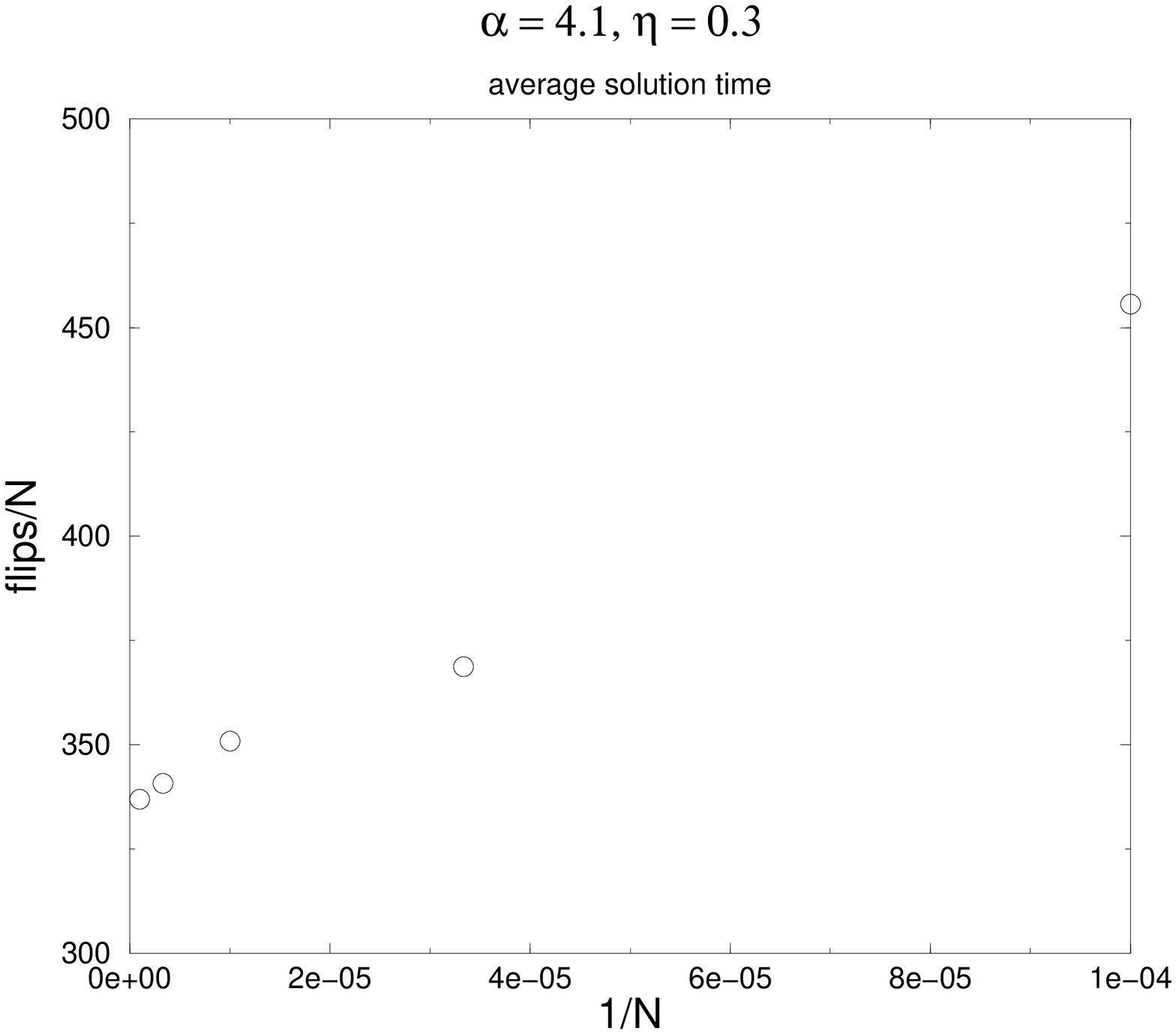,width=0.35\linewidth,angle=270}
}
\subfigure{
   \epsfig{file=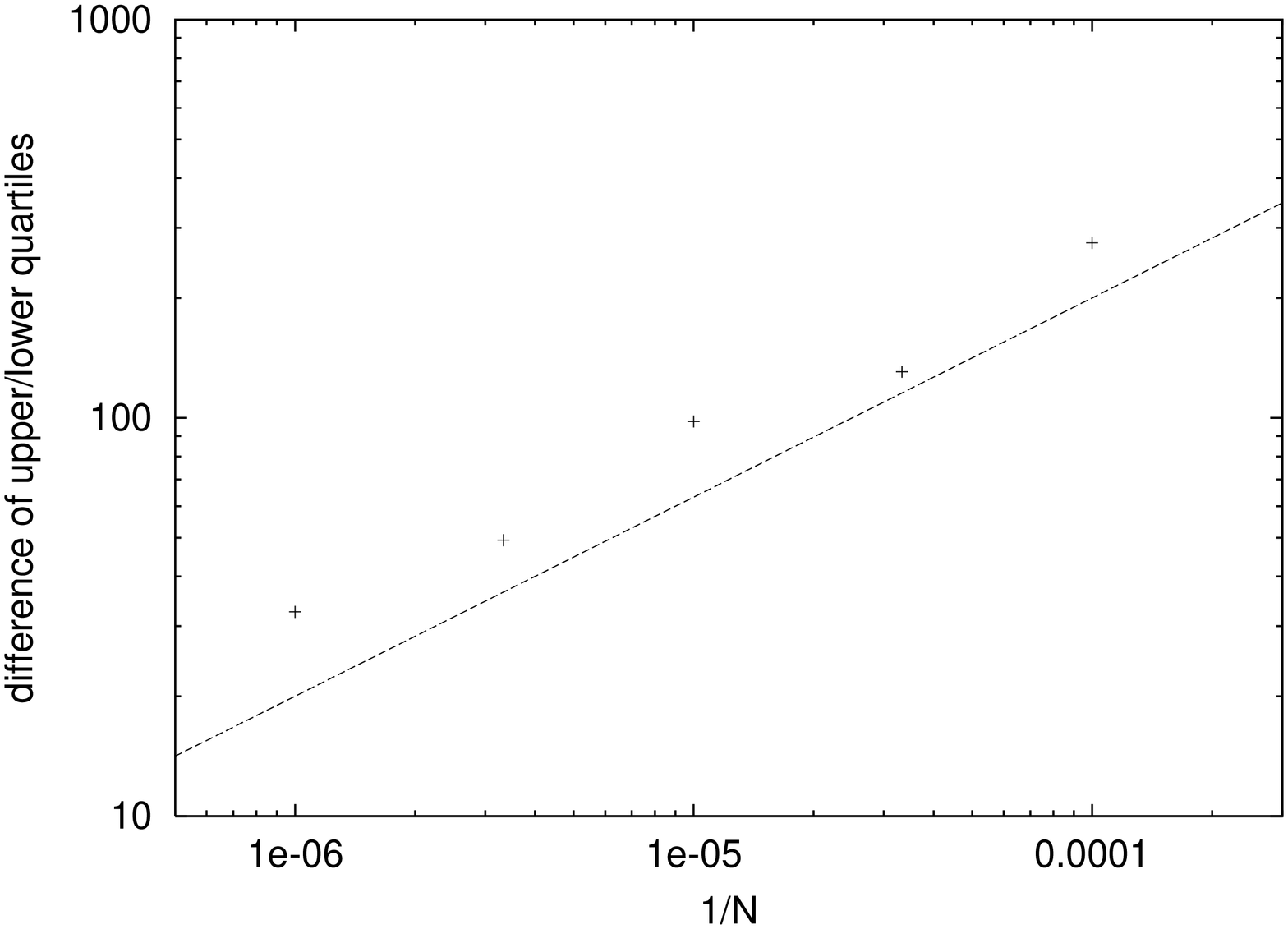,width=0.35\linewidth,angle=270}
}
\caption{
(a) The $N$-dependence of the average solution time for
$\alpha=4.1$ and $\eta=0.3$.
(b) The difference of the upper and lower quartiles
in $t_{sol}$ vs.\ $N$ for the same parameters.
The line shows $1/\sqrt{N}$-behaviour as a guide for the
eye.}
\label{fig:probdist}
\end{figure}

The linearity of FMS can fail due to the formation of heavy tails.
This, with a given $\alpha$ and a not too optimal, large $\eta$ would 
imply that the solution time $t_{sol}$ has at least a divergent mean
(first moment) and perhaps a divergent median as well. This can be 
deliberated upon by considering the "scaling ansatz"
\begin{equation}
P(t_{sol}) \sim (t_{sol})^{-a} f(t_{sol}/N^b)
\label{ansatz}
\end{equation}
where $f(x) = \mathrm{const}$ for $x$ small, and $f \rightarrow 0$
rapidly for $x \geq 1$. This simply states that for a fixed $N$ there
has to be a maximal solution time (even exponentially rare) since the
FMS is ``ergodic'' or able to get out of local minima.\footnote{This is true
for such problems where the focusing is not problematic; the XOR-SAT.for
instance can be different.} The condition that 
$\langle t_{sol} \rangle\sim N^{b(2-a)}$ 
be divergent with $N$ would then set a relation for the exponents
$a$ and $b$. Our experiments in the linear regime
have not yielded any evidence for a distribution following such
an ansatz, and moreover we have not explored systematically
the non-linear parameter region, where such behaviour might make
considering scalings as Eq.~(\ref{ansatz}) interesting. 
The average solution time, in terms of flips per spin,
is shown in Figure~\ref{fig:probdist}(a)  for $\alpha=4.1$.
Together with Figure~\ref{fig:probdist}(b), showing the tendency for the
width of the distribution to diminish as $1/\sqrt{N}$, this
$1/N$-behaviour implies rather trivial finite size corrections
to $P(t_{sol})$. In the Figure we depict the width of the distribution
$P$ measured by quantiles instead of the standard deviation,
since this is the most sensible measure given the nature
of the data. 

\begin{figure}\center
   \epsfig{file=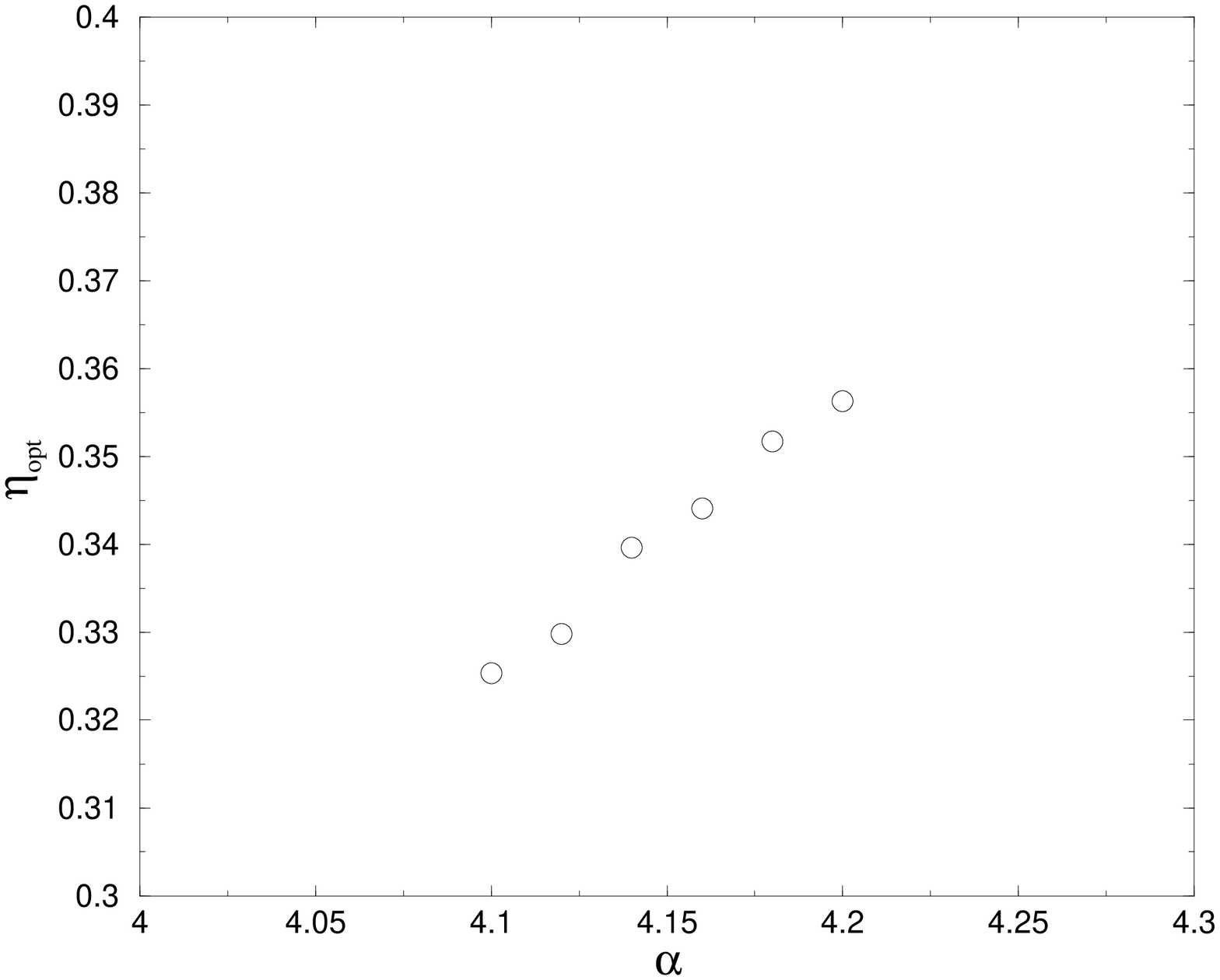,width=0.45\linewidth,angle=270}
\caption{Optimal $\eta$ vs. $\alpha$, for $N=10^5$.}
\label{fig:etaopt}
\end{figure}

\begin{figure}\center
   \epsfig{file=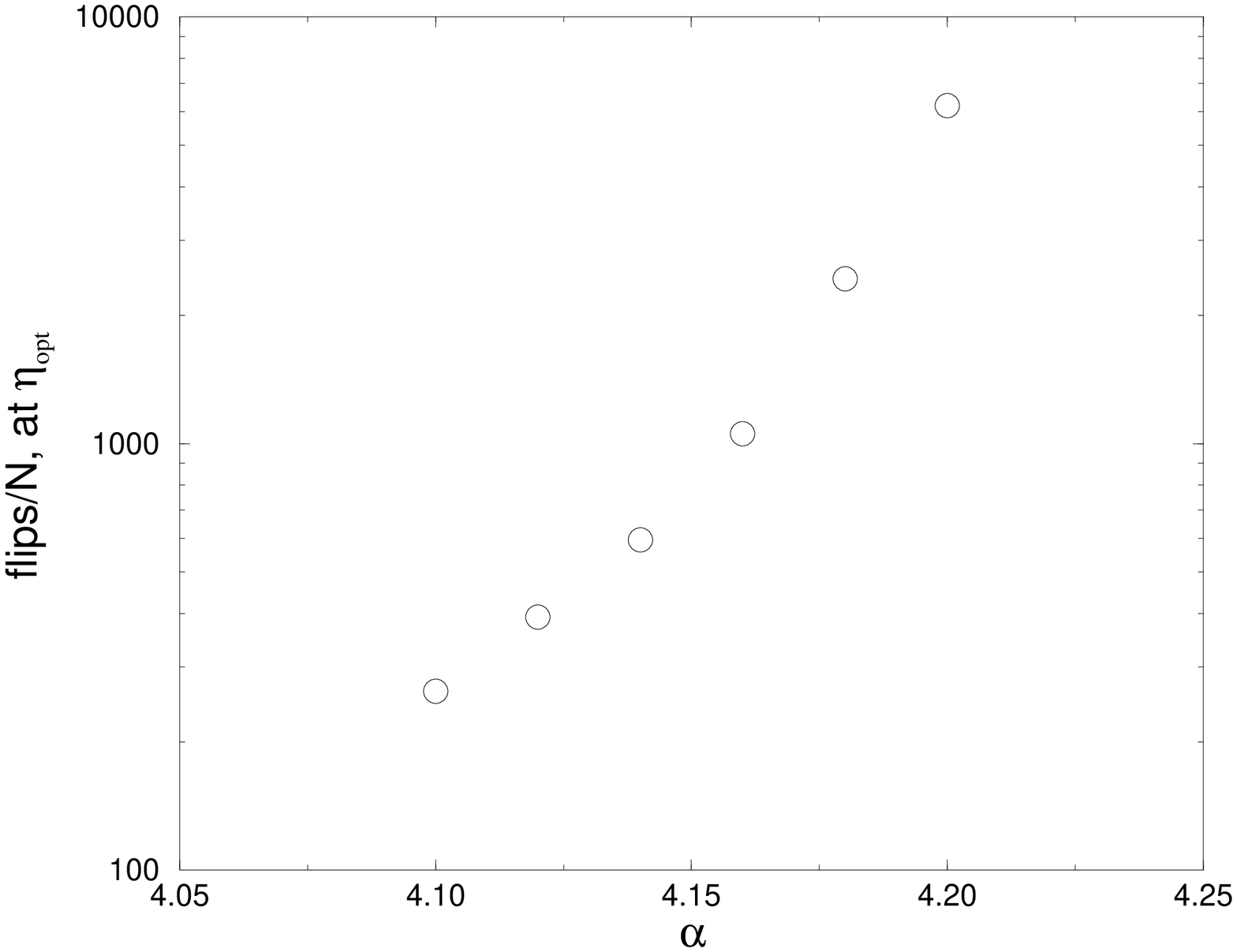,width=0.45\linewidth,angle=270}
\caption{Solution time at $\eta_{opt}$ vs. $\alpha$.}
\label{fig:etaoptime}
\end{figure}

\begin{figure}\center
   \epsfig{file=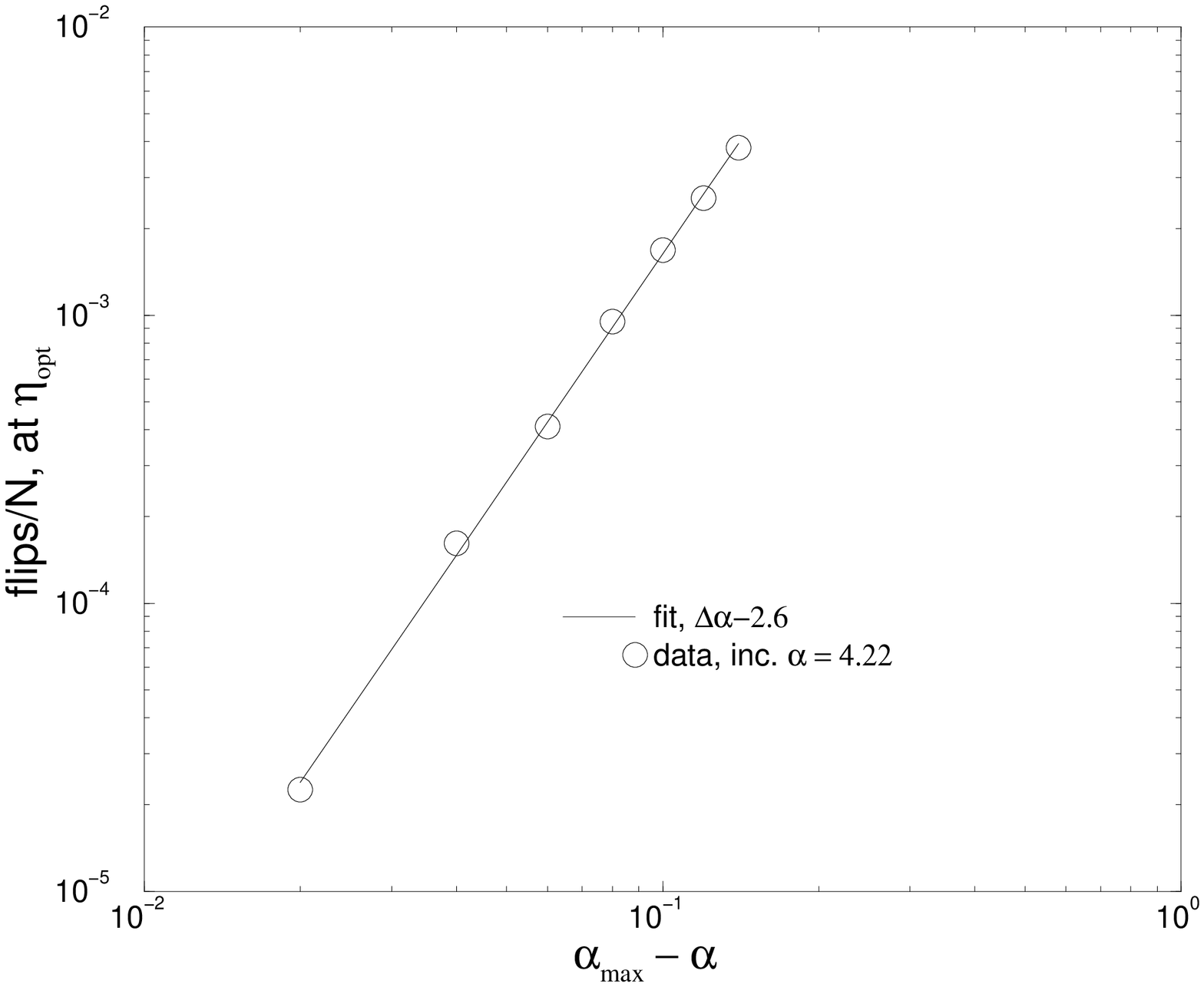,width=0.45\linewidth,angle=270}
\caption{Solution time $t_{sol}$ at $\eta = \eta_{opt}(\alpha)$
for $N=10^5$: possible divergence.}
\label{fig:topt}
\end{figure}

We also tried to extract the best possible performance
of the algorithm as a function of $\alpha$. Using the
data for varying $\eta$ allows one to extract roughly the
optimal values $\eta_{opt}(\alpha)$ which are demonstrated
in Figure~\ref{fig:etaopt}. As can be seen, the data
indicate a roughly linear increase of the optimal $\eta$
with in particular no notice
of the approach to $\alpha_c$ or to an algorithm-dependent
maximum value $\alpha_{max}$. The same data can be also utilised
to plot, for a fixed $N$ (recall the FMS runs linearly in
this regime) the solution time at the optimal noise
parameter $\eta$.
Figure~\ref{fig:etaoptime} shows that this as expected
diverges. Attempts to extract the value $\alpha_{max}$
limiting the linear regime by fitting
to various divergences of the kind $t_{sol} \sim (\alpha_{max}
- \alpha)^{-b}$ do not allow one to make a statistically reliable
conclusion as to whether $\alpha_{max} < \alpha_c$, though .
The reason for this is, as far as the data is concerned, the
distance of the $\alpha$ studied to any plausible value of $\alpha_{max}$.
See Figure~\ref{fig:topt}.

\section{Focused Record-to-Record Travel}

\begin{figure}\center
\begin{verbatim}
RRT(E,d):
  s = random initial configuration;
  s* = s; E* = E(s);
  while moves < max_moves do
    if s is a global min. of E then output s & halt,
    else:
      pick a random neighbour s' of s;
      if E(s') <= E* + d then let s = s';
      if E(s') < E* then:
        s* = s'; E* = E(s').
\end{verbatim}
\caption{The Record-to-Record Travel algorithm.}
\label{fig:rrt}
\end{figure}

Record-to-Record Travel,
a very simple stochastic local optimisation algorithm
outlined in Figure~\ref{fig:rrt}, was introduced
by Dueck in~\cite{Duec93}.
It restricts acceptable moves based on the best configuration
found so far, and is in this way non-Markovian.
Dueck claimed good results on solving 442-city
and 532-city instances of the Travelling Salesman Problem,
but after that the algorithm has been little used.
Initial experiments on applying the focused version
of this algorithm to the satisfiability problems were reported
in~\cite{SeOr03}.

In applying Record-to-Record Travel to satisfiability,
one again chooses as the objective function
$E(s) = $ number of clauses unsatisfied by
truth assignment $s$. The local search is performed 
over single-variable flip neighbourhoods.
We also impose the focusing condition requiring that the
flipped variables are always chosen from unsatisfied clauses.
(Precisely: one unsatisfied clause is chosen at random, and
from there a variable at random.) This leads to the
{\em focused} RRT (FRRT) algorithm.

\begin{figure}\center
\subfigure[Complete data]{
   \epsfig{file=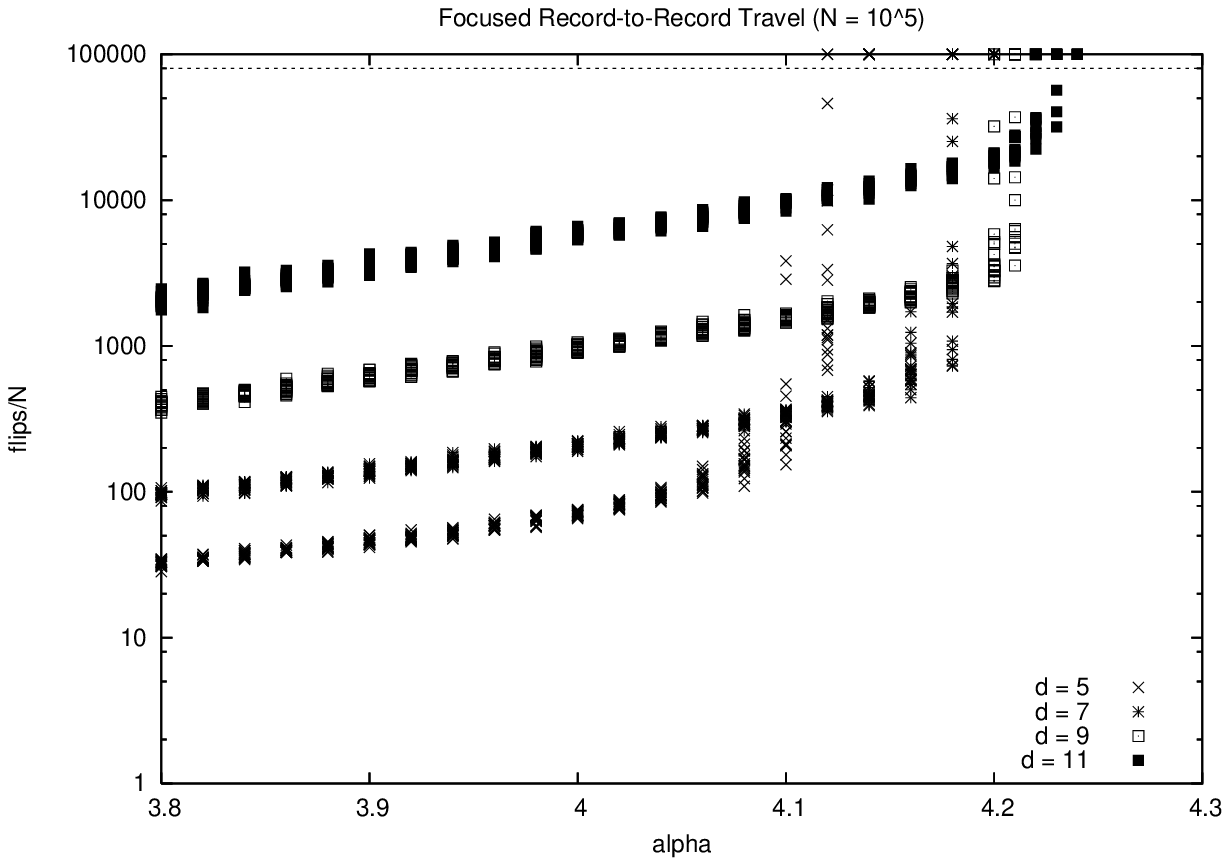,width=0.45\linewidth}
}
\subfigure[Medians and quartiles]{
   \epsfig{file=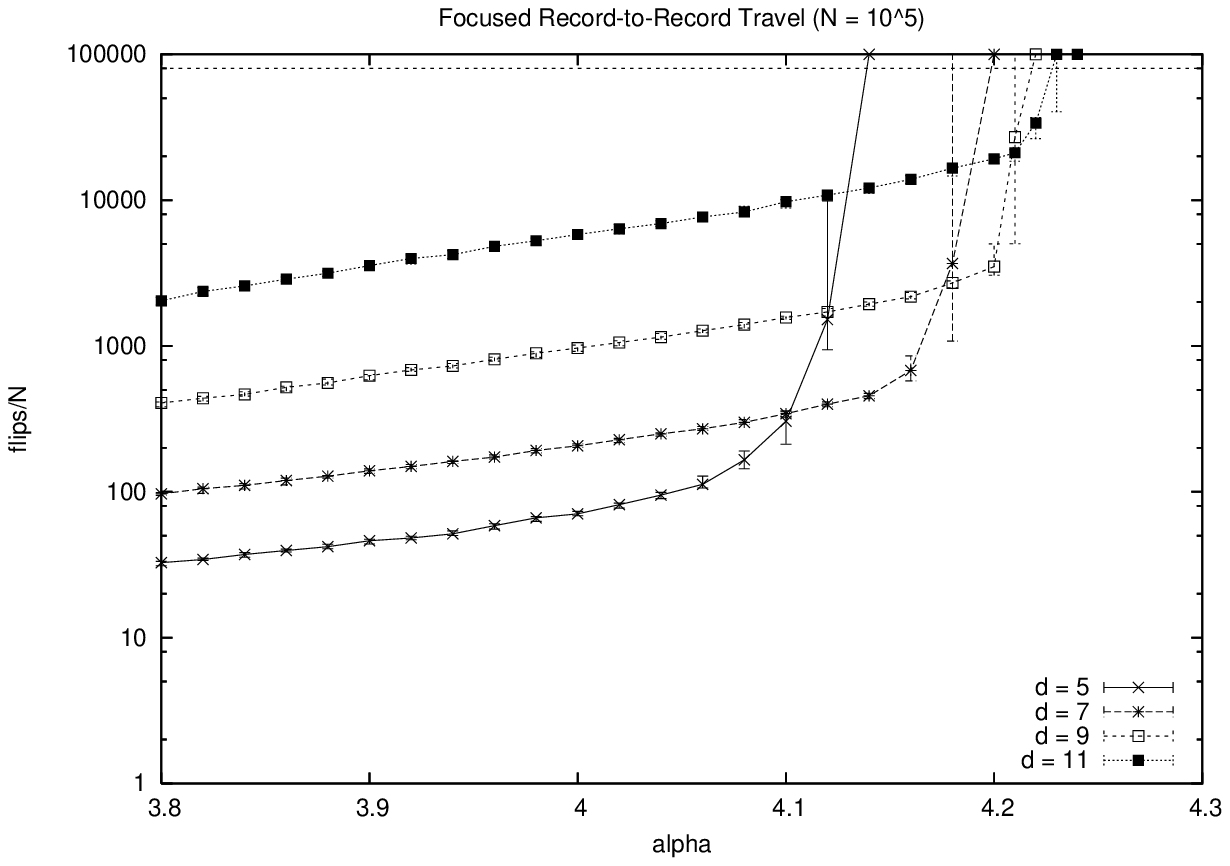,width=0.45\linewidth}
}
\caption{Normalised solution times for FRRT, $\alpha = 3.8\ldots 4.3$.}
\label{fig:frrt_scan}
\end{figure}

The algorithm is surprisingly powerful, as indicated
by the results in Figure~\ref{fig:frrt_scan}.
Interestingly, at least within the range of values
tested, the FRRT algorithm does not seem to have an
optimal value for the deviation parameter $d$, but
larger values of $d$ appear to always extend the linear
regime of the algorithm further, albeit with increasing
coefficients of linearity. E.g.\ for $d = 5$ the linear
regime seems to extend only to about $\alpha = 4.12$,
but for $d = 9$ already to $\alpha = 4.2$, and for
$d = 11$ even further.

\begin{figure}\center
\subfigure[$\alpha = 4.15$]{
   \epsfig{file=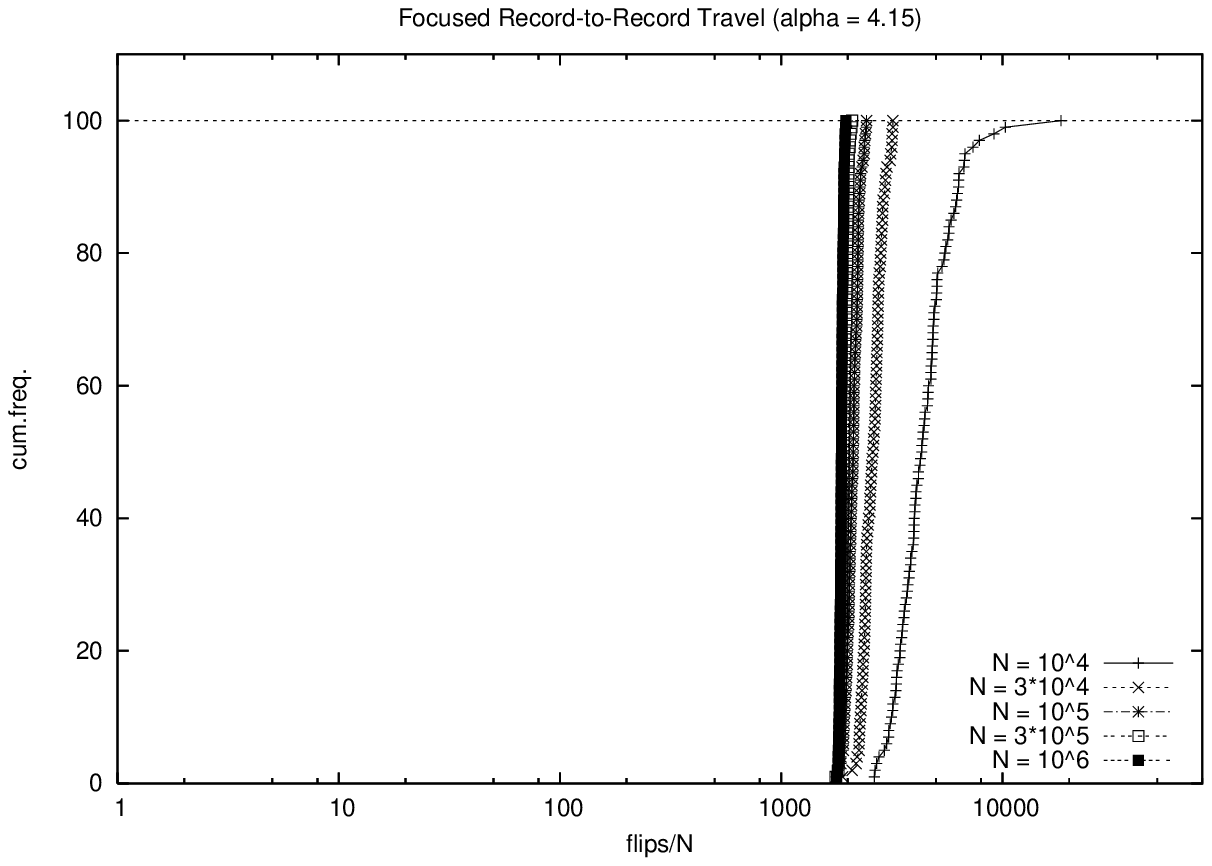,width=0.45\linewidth}
}
\subfigure[$\alpha = 4.20$]{
   \epsfig{file=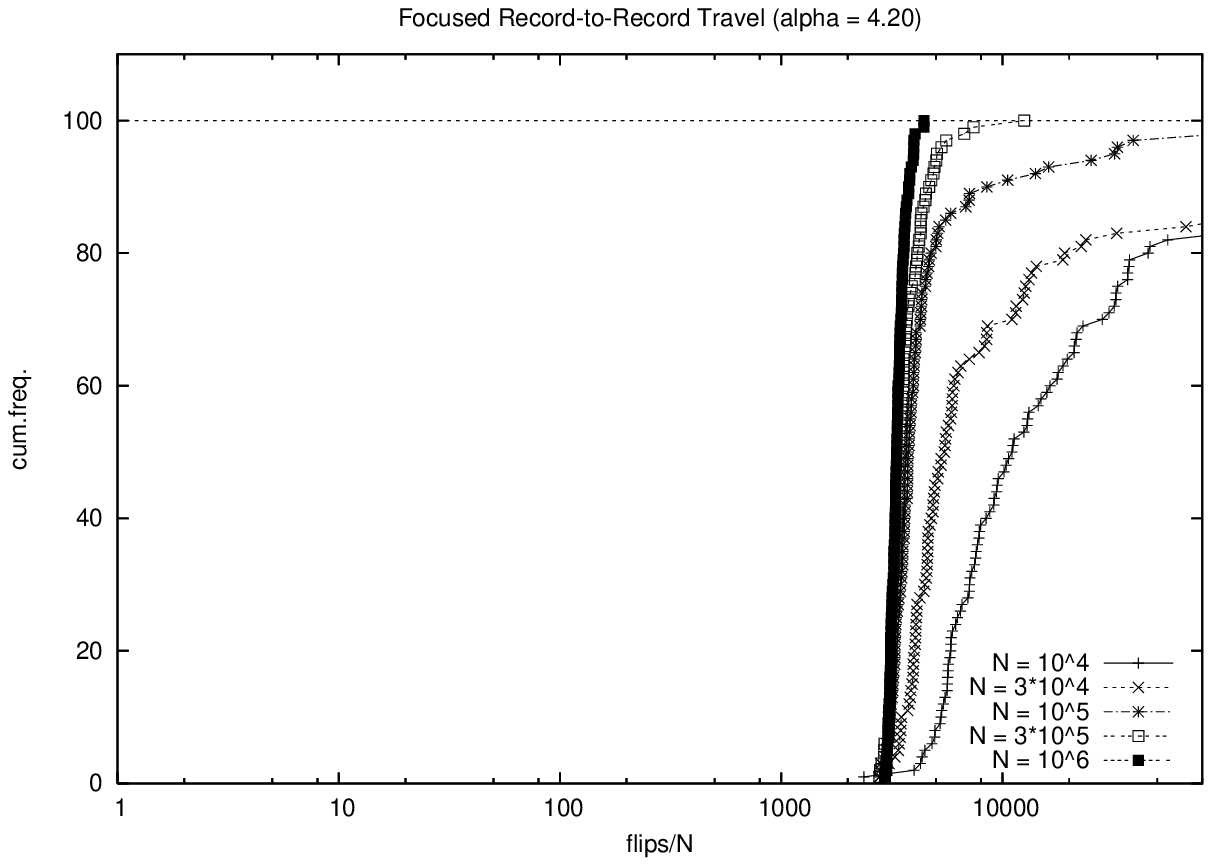,width=0.45\linewidth}
}
\caption{Cumulative solution time distributions for
FRRT with $d = 9$.}
\label{fig:frrt9_dist}
\end{figure}

To validate these results, we again determined the normalised
cumulative solution time distributions for the FRRT
algorithm with $d = 9$ at both $\alpha = 4.15$ and
$\alpha = 4.20$, similarly as in Figure~\ref{fig:wsat55_dist}.
As can be seen from Figure~\ref{fig:frrt9_dist}, FRRT
behaves for these parameter values more predictably
than WalkSAT at $p = 0.55$, and comparably to WalkSAT
at the near-optimal noise parameter value of $p = 0.57$.
(Interestingly, the FRRT normalised median solution times
seem to decrease slightly with increasing $N$. This phenomenon
is undoubtedly connected to the smoothing of the random 3-SAT
energy landscapes for increasing $N$, but otherwise we have no
clear explanation for it. It can also be observed that the 
shapes of the cumulative distributions are qualitatively different
from the FMS case.)

\begin{figure}\center
\subfigure[$\alpha = 4.15$]{
   \epsfig{file=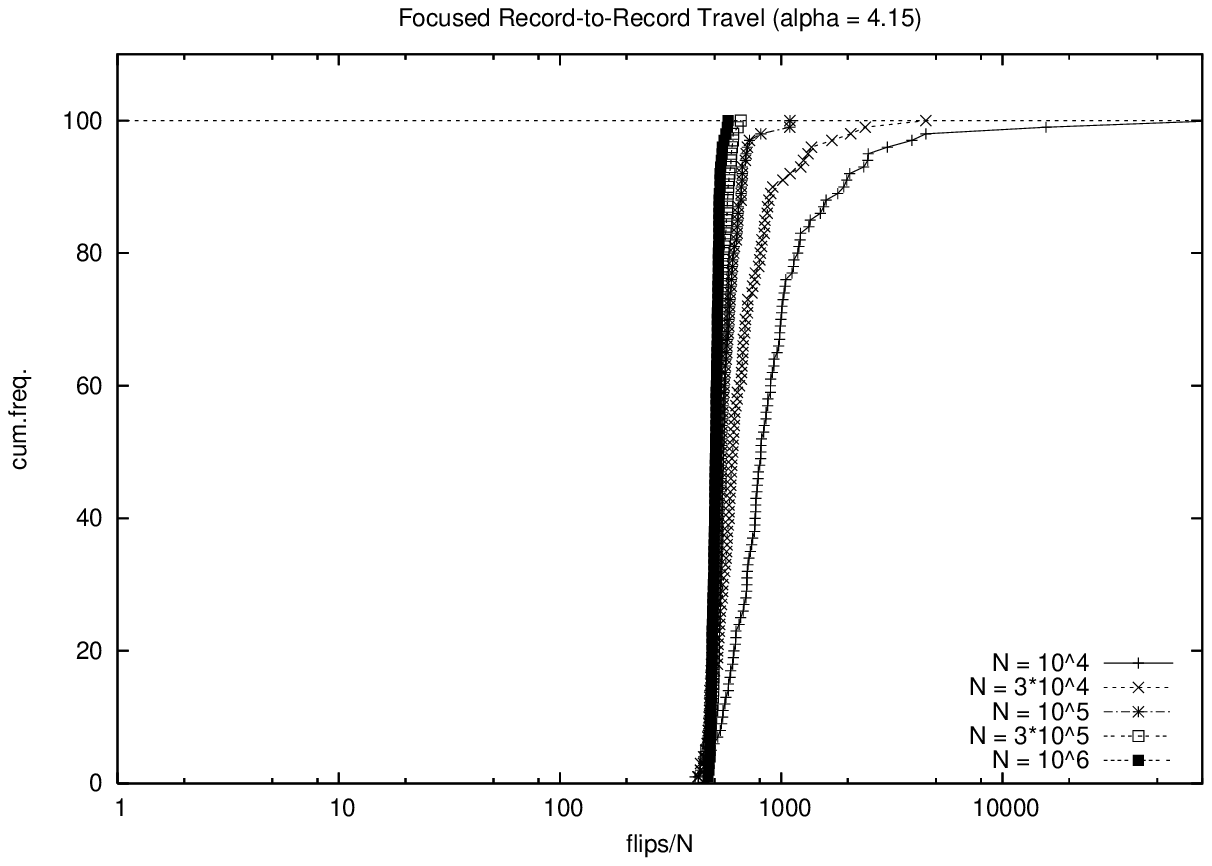,width=0.45\linewidth}
}
\subfigure[$\alpha = 4.20$]{
   \epsfig{file=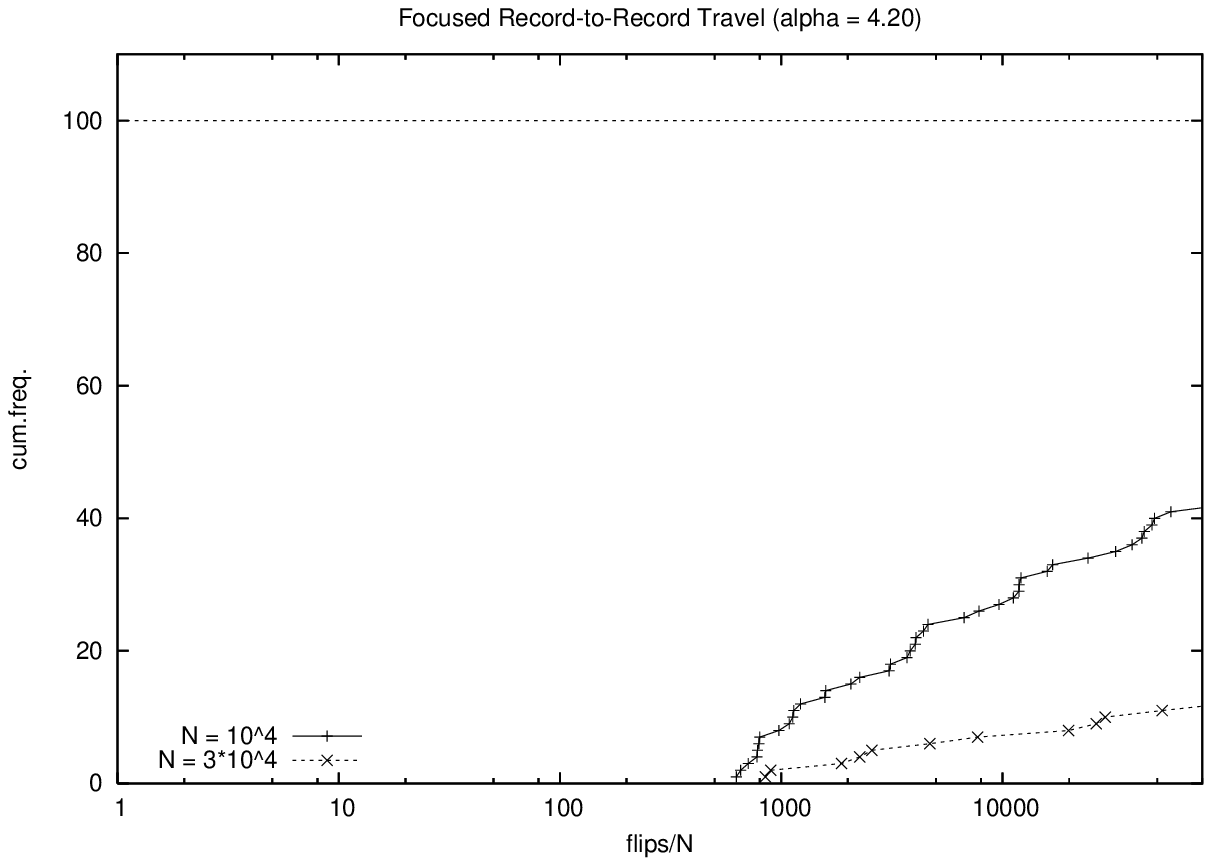,width=0.45\linewidth}
}
\caption{Cumulative solution time distributions for
FRRT with $d = 7$.}
\label{fig:frrt7_dist}
\end{figure}

For comparison, Figure~\ref{fig:frrt7_dist} illustrates
the corresponding distributions for $d = 7$.
As can be seen, for $\alpha = 4.15$ FRRT converges even
with $d = 7$, but for $\alpha = 4.20$ a larger value
of $d$ is required.

\section{Whitening and the Linear Time Regime}

As already mentioned in Section 2, WalkSAT finds a solution in linear 
time when $\alpha < \alphadyn(p)$, where $\alphadyn$ is a dynamical 
threshold depending on the noise parameter $p$. When 
$\alpha > \alphadyn$, the energy falls to a metastable level around 
which it remains fluctuating. In this metastable state new unsat 
clauses are produced at approximately the same rate as 
they are being eliminated, and a solution is found only by a lucky 
fluctuation, for which the waiting time is exponential in 
$N$~\cite{BaHW03, SeMo03}. 

Our experiments show this kind of behaviour also for FMS.\footnote{
FRRT, on the other hand, does not seem to suffer from this 
kind of metastability, because every time a new ``record'' is found, 
the ceiling for the allowed energies is lowered. This partially 
explains why increasing the deviation parameter $d$ seems to
always extend the linear time regime further.} When the
parameter $\eta$ is decreased away from one, the corresponding
$\alphadyn(\eta)$ value increases (cf.\ Figure~\ref{phase} in Section 6).
Since the rate of production of unsatisfied clauses vanishes
as $\eta$ decreases toward zero, it is tempting to think that
if FMS is in the linear time regime for some particular values
of $\alpha$ and $\eta$, then 
it remains so also for all smaller $\eta$ given that particular 
$\alpha$. But this seems not always to be the case. 
 
It is a common observation that excessive greediness can be detrimental
to the quality of optimisation results. How is such behaviour
exhibited in the case of the FMS algorithm on random 3-SAT?
We think that the answer is related to the 
concept of {\em whitening}, introduced in the context of the graph
colouring problem (hence the name) in \cite{Pari02} and applied to
random 3-SAT in \cite{BrZ03, MaMW04}. The constraints 
between variables contain slack that can be consumed too quickly by a 
greedy search. This slack is due to ``safe'' clauses, i.e.\ clauses 
that contain at least two true literals.
For some $\alpha$ values, if $\eta$ is too low, 
FMS typically drives the system to a state
where some unsatisfied clauses can no longer be eliminated
``locally''.\footnote{In the limiting case of an infinite tree,
a ``local'' subtree is such that its longest path from the root
has finite expected length.}

\begin{figure}\center
\begin{verbatim}
WHITENING(F, s):
  mark all the clauses white (*, joker, wild card) except those that 
    have only one true literal (thus, also unsatisfied clauses are marked 
    white, so that they do not affect the whitening process);
  loop
    - mark those variables white that appear as satisfying literals 
      only in white clauses;
    - halt, if all the variables are white (s is completely white);
    - halt, if no new variables became white in this round (s has a core);
    - mark those clauses white that contain at least one white variable.
\end{verbatim}
\caption{The whitening algorithm for a truth assignment s.}
\label{fig:whitening}
\end{figure}

The {\em whitening algorithm}\footnote{We have extended the concept
of  whitening so that it can be applied also to truth assignments 
that are not satisfying solutions. All the unsat clauses 
are just ignored, making the truth assignment a solution for 
the remaining clauses.} presented in Figure~\ref{fig:whitening}
searches for {\em frozen} variables in a given truth assignment. 
Flipping any frozen variable produces at 
least one new unsat clause that contains only frozen variables. 
To get rid of that unsat clause, one of its variables must be 
flipped, and that produces again at least one new unsat clause 
with all of its other variables frozen, and so on. The only way to 
do away with all of these newly produced unsat clauses is by 
loops, i.e.\ by flip sequences that encounter some variables more than 
once. In random 3-SAT instances these loops are typically long, growing as 
$O(\log N)$. So when $N \to \infty$, also the loops become infinitely long. 
If there are no frozen variables, the truth assignment is said to be 
{\em completely white},
otherwise it has a {\em core} of frozen variables (the term 
core in this connection is from \cite{MaMW04}).

When energy decreases during a search, a transition can
occur from a state where the encountered truth assignments
are generically
completely white to a state where the truth assignments
generically have a core of frozen variables.
We call this the {\em freezing transition}. It is natural 
to assume that when a freezing transition has happened, the remaining 
unsatisfied clauses cannot be eliminated in linear time, because then 
the search needs loops growing with the system size.

\begin{figure}\center
   \epsfig{file=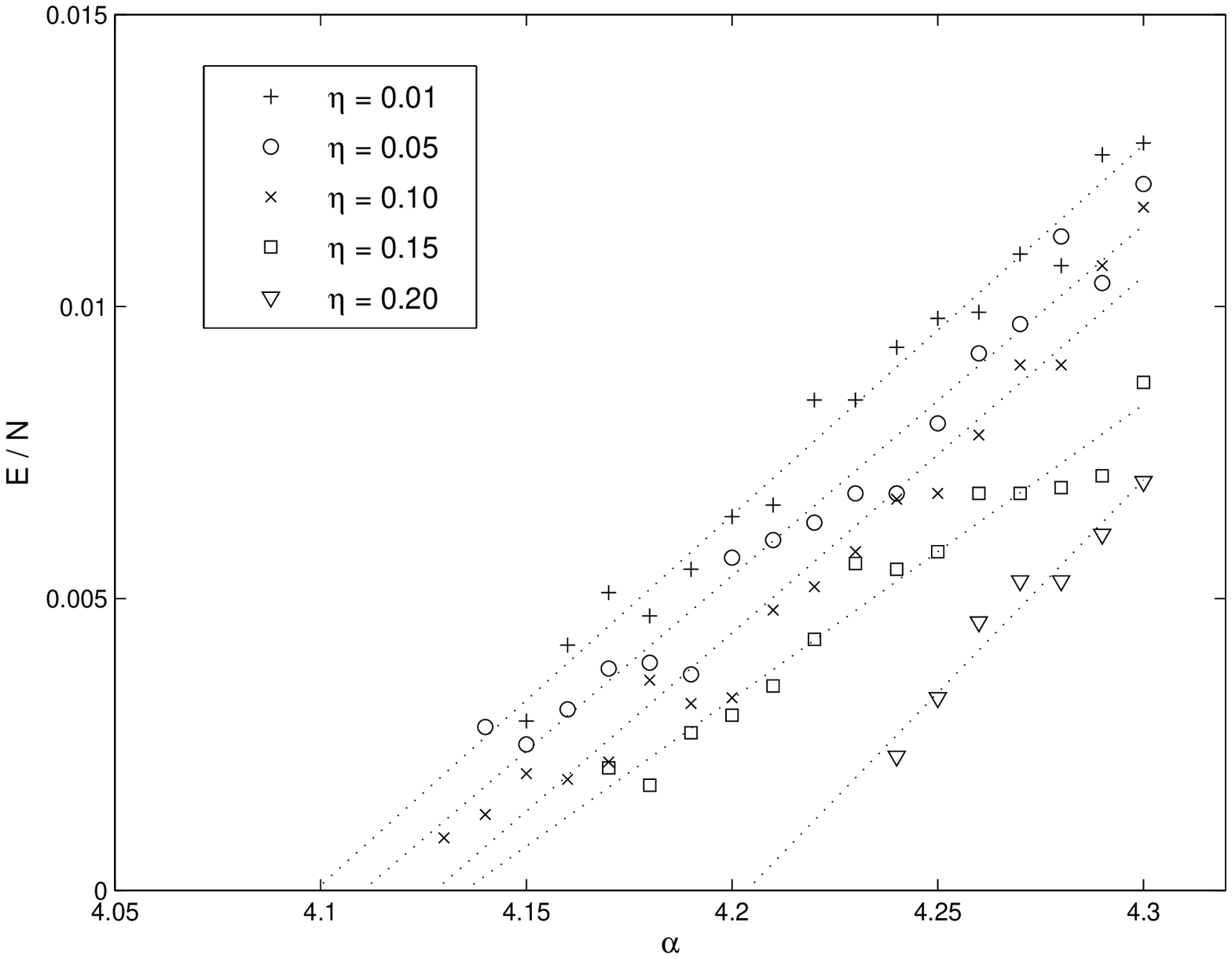, width=0.55\linewidth}
\caption{Freezing transition energies $e_f$, when using FMS.
Each point corresponds to a single run ($N = 10^5$).
Dotted lines are linear fits.}
\label{fig:frtransit}
\end{figure}

\begin{figure}\center
   \epsfig{file=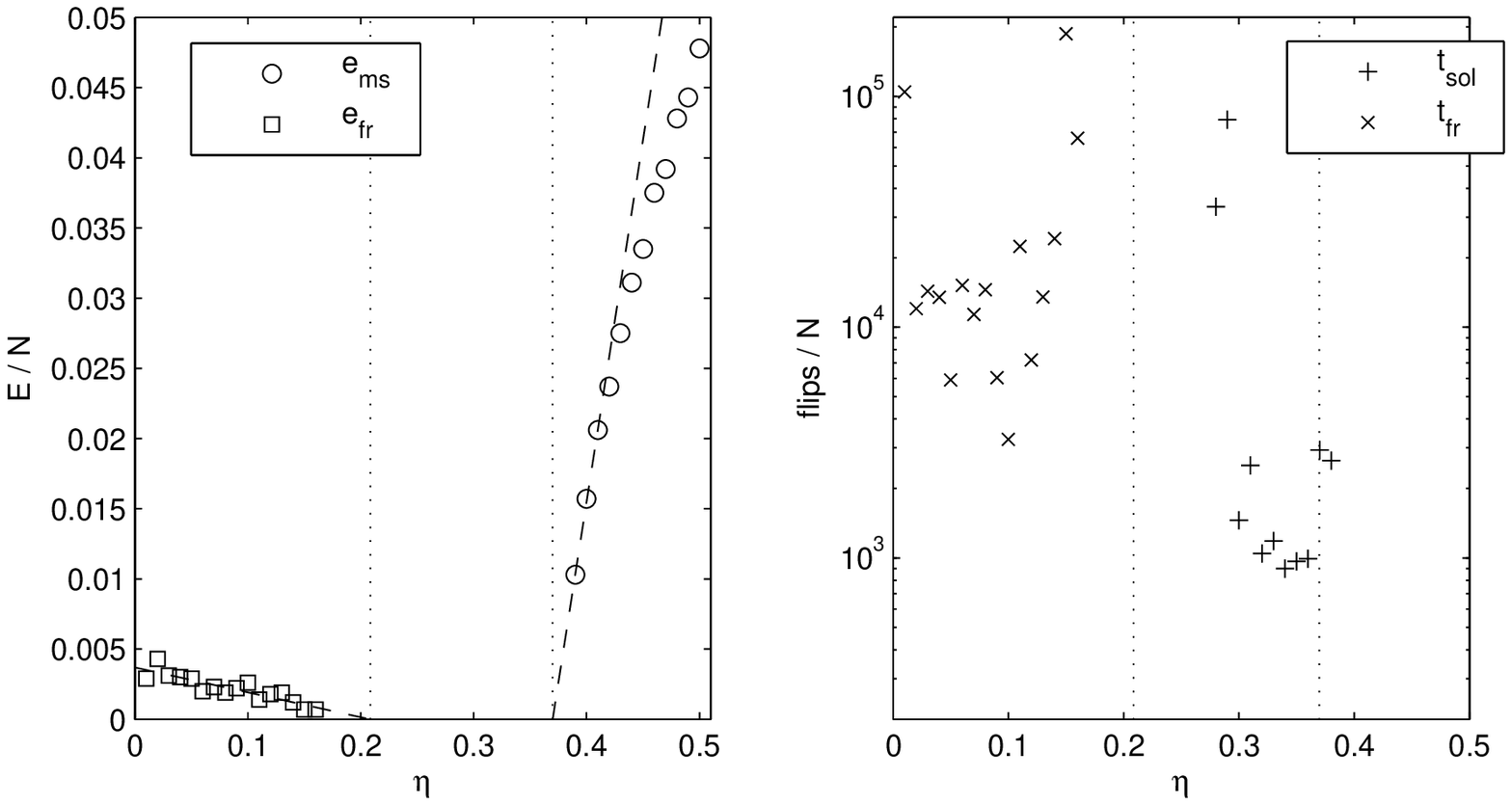, width=0.75\linewidth}
\caption{Sample runs of FMS at $\alpha = 4.15$ ($N = 10^5$).
Left: Metastable energies and freezing transition energies.
Right: Solution times and freezing transition times.
Dotted lines give a rough estimate of the $(\eta_l, \eta_u)$-window
at this $\alpha$.}
\label{fig:alpha415}
\end{figure}

\begin{figure}\center
   \epsfig{file=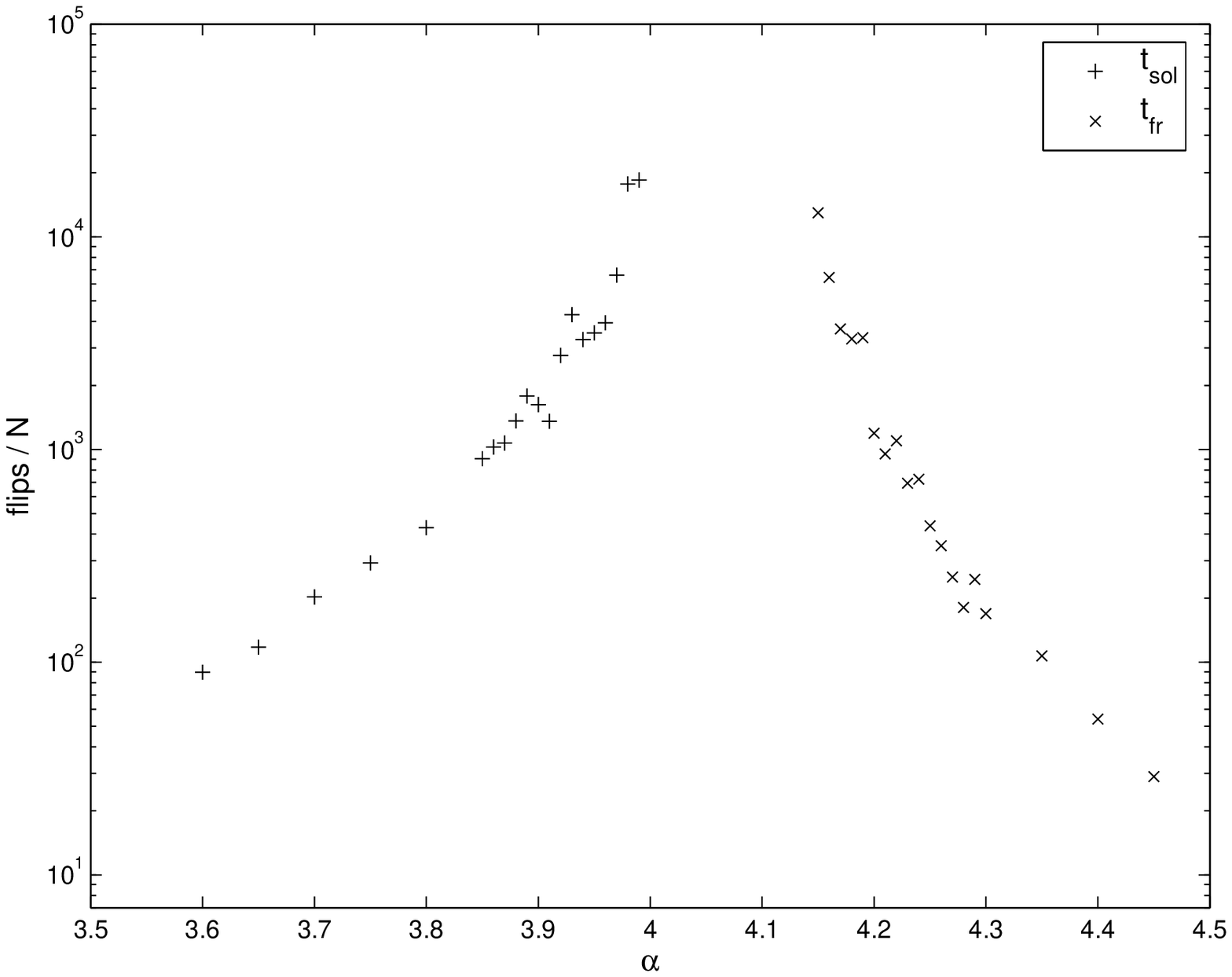, width=0.55\linewidth}
\caption{Solution times and freezing times, when using FMS 
with $\eta = 0.05$.
Each point corresponds to a single run ($N = 10^5$).}
\label{fig:eta05}
\end{figure}

In Figure~\ref{fig:frtransit} we present estimates of the freezing
transition energy density $e_f \equiv E_f / N$ for a few sample
runs of the FMS algorithm.
We have not studied the dependence of $E_f$ on N, but it appears to
be self-averaging, so that for any $\alpha$ and $\eta$ for which
a freezing transition occurs,
$e_f$ converges to a constant value as $N \to \infty$.
Hence the $e_f$ values given in Figures \ref{fig:frtransit} and 
\ref{fig:alpha415} are similar to their $N \to \infty$ limits.
Also the freezing transition time $t_f$ (measured in flips/$N$) seems
to converge for many $(\alpha,\eta)$-pairs, but it is not at all 
clear that this needs to be so in every case. There may exist some 
region of $\alpha$ and $\eta$ values for which a superlinear
number of flips are required before either a satisfying solution
or a truth assignment with a core is reached (cf.\ Figure~\ref{fig:eta05}).

We conjecture that there are almost always truth assignments with a core 
at all energy levels below any level at which a freezing transition 
can happen at given $\alpha$. For example, this implies that one can pick
an $\eta$-value for $\alpha=4.2$, such that the freezing takes place 
at, say, $E/N=0.005$ (Figure \ref{fig:frtransit}).
Hence we believe that there exist solutions 
with a core also in the $N \to \infty$ limit in some region of 
$\alpha$-values, approximately $4.10 \lesssim \alpha < \alpha_c$.
The lower bound of 4.10 has been estimated using FMS
with $\eta = 0.01$ (Figure \ref{fig:frtransit}).
Since the choice of algorithm may have an effect on the estimate,
the actual threshold point below which there do not exist solutions
with a core in the $N \to \infty$ limit can be lower than this.

There may be a narrow region below $\alpha_c$ where 
typical instances do not have any completely white solutions.
In this region, if it exists, focused algorithms presumably 
cannot find a solution in linear time.
On the other hand, in~\cite{MaMW04} it is conjectured that almost
all solutions are completely white
(``all-$*$ assignments'' in the terminology of~\cite{MaMW04})
up to $\alpha_c$.
This claim is based on experimental evidence using the survey propagation 
algorithm~\cite{BrMZ02, MeZe02, BrZ03}, which seemingly finds 
only completely white solutions when N is large~\cite{MaMW04}, 
just like focused local algorithms.
However, if solutions with a core cannot be found in linear time,
it is not surprising that usually the first solution encountered
is completely white. Thus, the issue of a possible core transition
(similar to the simpler XORSAT problem~\cite{MeRZ03}) 
below $\alpha_c$ remains open.
We have not tried to assess which type of solution is more 
numerous near $\alpha_c$, or whether having
a majority of completely white solutions
is in fact a necessary or even a sufficient condition
for maintaining time-linearity of local search methods.
Thus, we can at the moment only conjecture a relation
between the algorithmic performance of local search methods and
the entropy of the solution space, as measured in terms of
the ``white'' and ``core'' fractions vs.\ $N$.

\begin{figure}\center
   \epsfig{file=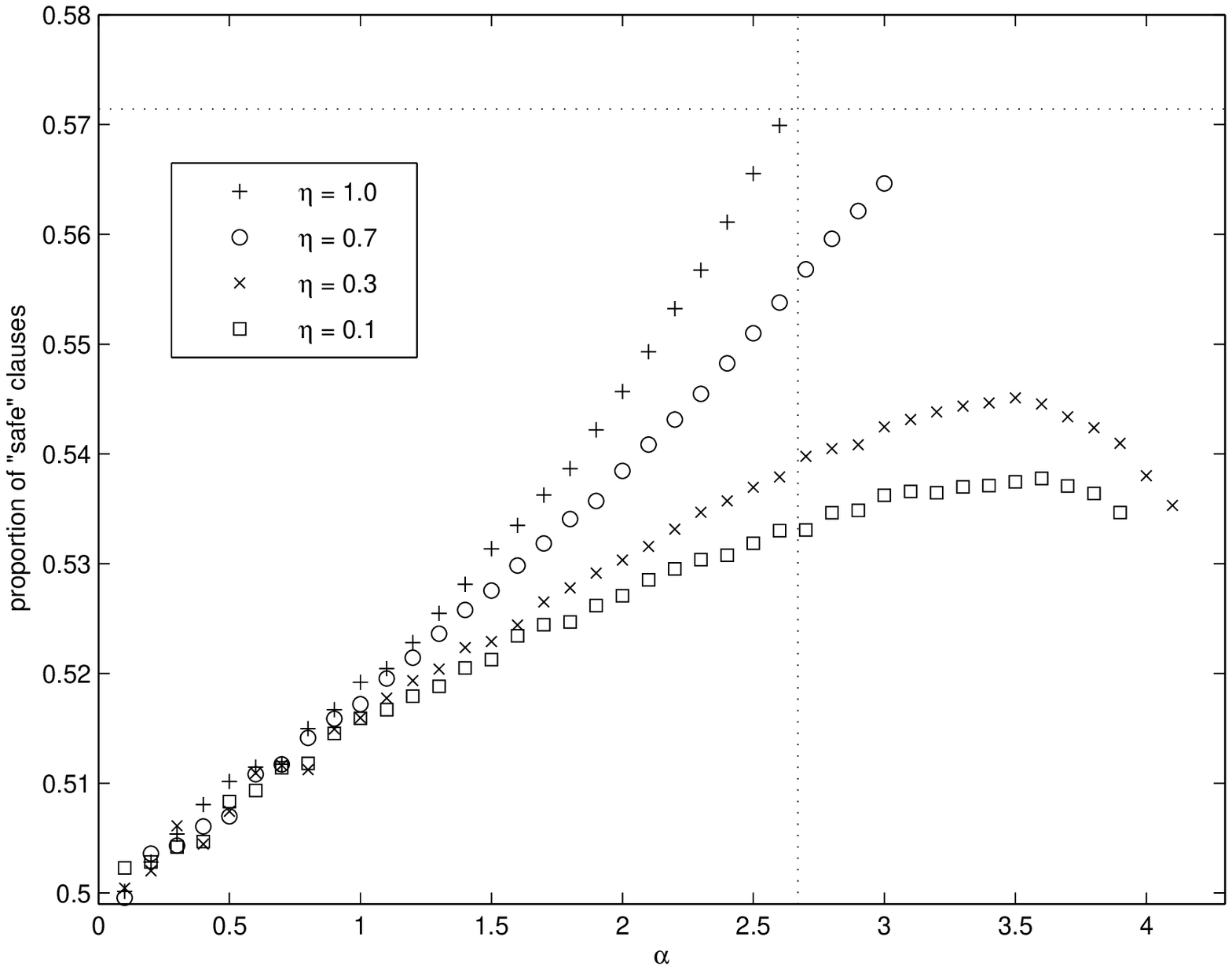, width=0.55\linewidth}
\caption{Proportion of ``safe'' clauses in solutions found by FMS.
Each point corresponds to a single run ($N = 10^6$).
Horizontal dotted line is this proportion, when all the satisfying 
combinations have equal probability: $4/7$.
Vertical dotted line is our estimate for the dynamical threshold
of the Random Walk: $\alpha \approx 2.67$. }
\label{fig:safeprop}
\end{figure}

Another way to establish the existence of qualitatively different
kinds of solutions is to look at the proportion of ``safe'' clauses
(i.e.\ clauses that contain at least two satisfying literals)
in the solutions found by FMS with different values of $\eta$
(Figure~\ref{fig:safeprop}).
One can say that with greater $\eta$, better quality solutions are found.
Note that these safe clauses are just what enable variables to
become unfrozen in the whitening algorithm (Figure~\ref{fig:whitening}).

Whiteness is just one indicator that can be used during the search 
process to tell when the linear-time property has been lost. Certainly 
there can be more powerful indicators that alarm about difficulties 
earlier, when the freezing transition has not yet happened. If a 
truth assignment which is not a solution is completely white, 
it means that any {\em one} of the still remaining unsat
clauses can be resolved locally (assuming dependency loops
between clauses don't incidentally block this).
It does not, however, mean that an arbitrary subset of the 
remaining unsat clauses could necessarily be resolved locally, 
or that there necessarily exists a solution. A truth assignment can 
turn from a completely white one to having a core by the flip of 
just one variable. One can envision higher order whitenings: a $k$th 
order whitening would tell whether {\em all} truth assigments that
differ in at most $k$ variable values from the assignment in question are
completely white (in the usual sense, i.e.\ zeroth order whitened). 
Maybe a higher order whitening with any fixed $k$ can
(in principle\footnote{Note that it takes $O(N^k)$ normal whitenings 
to compute one $k$th order whitening, if done in a simple 
``brute force'' manner. One normal whitening 
can be done in time $O(N)$, if implemented well.}) be used as 
an indicator for the linear-time property,
just like the zeroth order one. There can also be some indicators 
that have nothing to do with whitening.\footnote{Like getting too 
close to a local optimum, which can hamper FRRT when using a small 
value of $d$~\cite{SeOr03}.}

\section{Conclusions}

In this paper we have elucidated the behaviour of ``focused'' local search
algorithms for the random 3-SAT problem. An expected conclusion is
that they can be tuned so as to extend the regime of ``good'',
linear-in-N behaviour closer to $\alpha_c$. 

\begin{figure}\center
   \epsfig{file=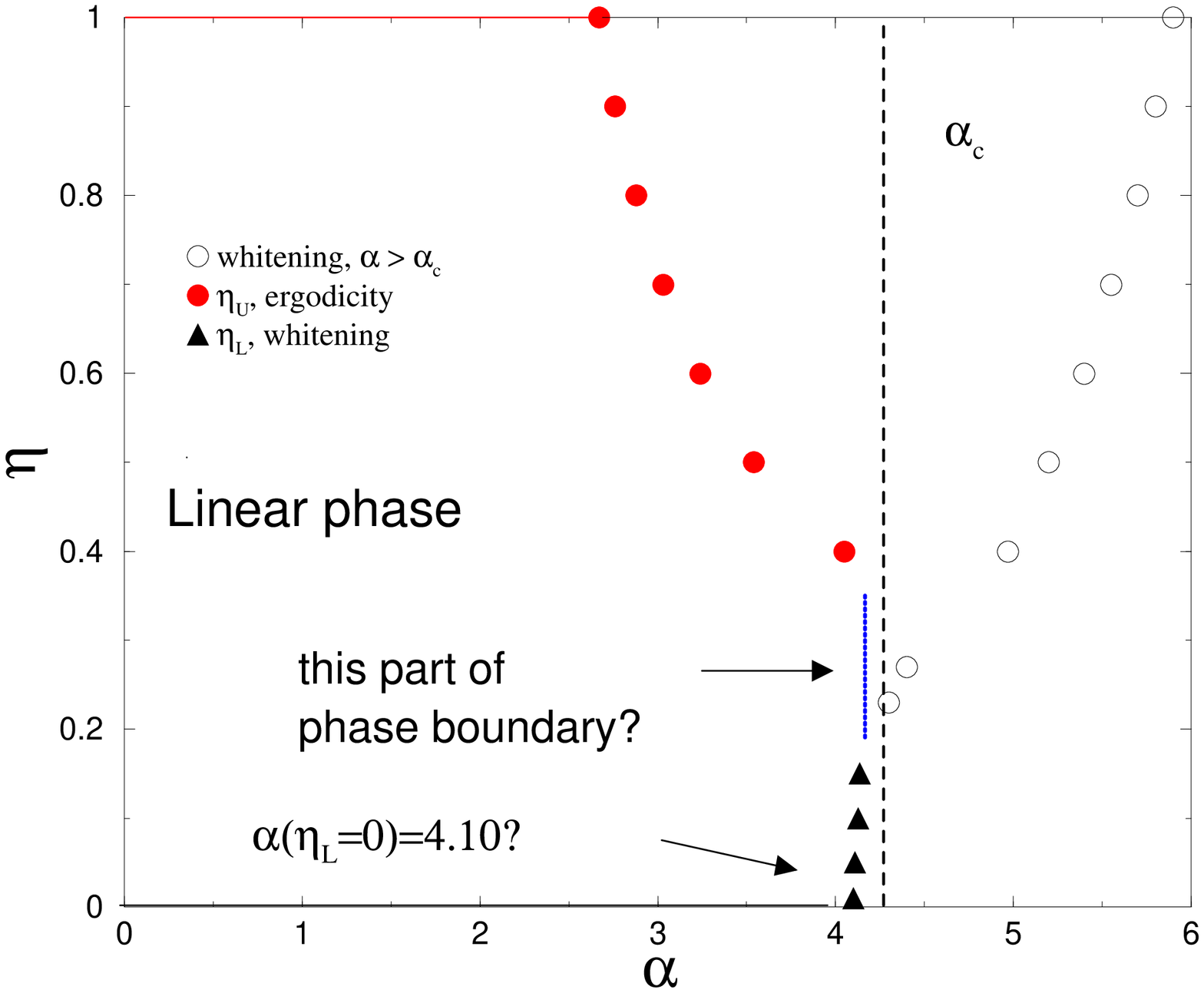,width=0.45\linewidth,angle=270}
\caption{Proposed phase diagram for FMS.
Above a dynamical threshold a metastable state is encountered
before a solution is reached. Below a whitening threshold
a freezing transition happens before a solution or a metastable or stable
state is reached. 
The data points are rough estimates based on a few runs of the FMS
algorithm.}
\label{phase}
\end{figure}

An unexpected conclusion is that FMS (and WalkSAT, and FRRT)
seem to work well very close to the critical threshold.
Figure~\ref{phase} proposes a phase diagram.
FMS with $\eta = 1$ is just the Random Walk algorithm;
hence the first transition point is at $\alpha \approx 2.67$.
For larger $\alpha$ there is the possibility of having
{\em two phase boundaries} in terms of the noise parameter. The upper
value $\eta_u$ separates the linear regime from one with too much noise,
in which the fluctuations of the algorithm degrade its performance.
For $\eta < \eta_l$, greediness leads to dynamical freezing,
and though FMS (in contrast to FRRT in particular) remains ergodic,
or able to climb out of local minima,
the algorithm no longer scales linearly in the number of variables.

The phase diagram presents us with two main questions: what is the 
smallest $\alpha$ at which $\eta_l(\alpha)$ starts to deviate from
zero? Does the choice of an ideal $\eta_{opt}$ allow one to push the
linear regime arbitrarily close to $\alpha_c$? Note that the deviation
of $\eta_u(\alpha)$ from unity with increasing $\alpha$
could perhaps be analysed with similar 
techniques as the methods in refs.~\cite{BaHW03,SeMo03}. The essential
idea there is to construct rate equations for densities of variables
or clauses while ignoring correlations of neighboring ones. This
would not resolve the above two main ones, of course.

The resolution of these questions will depend on our understanding
the performance of FMS or other local search 
presence of the clustering of solutions. The whitening or freezing
experiments provide qualitative understanding on the behaviour for
small $\eta$: too greedy search procedures are not a good idea.
Thus FMS seems to work well only when it actually finds solutions
that can be completely whitened. 
Figure~\ref{phase} also points out,
via the whitening data for $\alpha>\alpha_c$, that
the line of whitening thresholds $\eta_l$ for FMS apparently is
continuous over $\alpha_c$.
In particular in this case, one can then separate between two
possible scenarios: either the $\eta_u$-line extends to $\alpha_c$
and finishes there, or otherwise it meets with the $\eta_l$-line
at an $\alpha<\alpha_c$ at a point that is ``tri-critical'' in
statistical mechanics terms. Note, however, that $\eta_l$
determined by whitening is not necessarily the highest lower bound
to the linear phase.

The range of the algorithm presents therefore us with a fundamental
question: though replica-methods have revealed the presence of clustering
in the solution space starting from $\alpha \approx 3.92$, the FMS works
for much higher $\alpha$'s still. The clustered solutions
should have an extensive core of frozen variables, and 
therefore to be hard to find.
Thus, the existence of ``easy solutions'' for the FMS tells that there
are features in the solution space and energy landscape that are not well
understood, since the FMS can pick, for a right choice of $\eta$ {\em almost
always} such easy solutions out of those available for any particular
instance. This would seem to imply that such completely white ones
have a considerable entropy. 

Our numerics also implies that in the linear scaling regime the solution
time probability distributions become sharper (``concentrate'') as $N$
increases, implying indeed that the median solution time scales linearly.
We can by numerical experiments not prove that this holds also for the
average behaviour, but would like to note that the empirically observed
tail behaviours for the same distributions give no indication of such
heavy tails that would contradict this idea.

{\bf Acknowledgements:} research supported by the Academy of Finland, Grants 206235
(S. Seitz) and 204156 (P. Orponen), and by the Center of
Excellence program of the A. of F. (MA, SS) The authors would
like to thank Dr. Supriya Krishnamurthy for useful comments and
discussions.


\begin{thebibliography}{99}

\bibitem{AaLe97} E. Aarts and J. K. Lenstra (Eds.),
{\it Local Search for Combinatorial Optimization}.
J. Wiley \& Sons, New York NY, 1997.

\bibitem{AuGK04} E. Aurell, U. Gordon and S. Kirkpatrick,
Comparing beliefs, surveys and random walks.
Technical report cond-mat/0406217, arXiv.org (June 2004).

\bibitem{BaHW03} W. Barthel, A. K. Hartmann and M. Weigt,
Solving satisfiability problems by fluctuations: The dynamics
of stochastic local search algorithms.
{\it Phys. Rev. E 67} (2003), 066104.

\bibitem{BrMZ02} A. Braunstein, M. M\'ezard and R. Zecchina,
Survey propagation: an algorithm for satisfiability.
Technical report cs.CC/0212002, arXiv.org (Dec.\ 2002).

\bibitem{BrZ03}A. Braunstein and R. Zecchina, 
Survey propagation as local equilibrium equations,
Technical report cond-mat/0312483, arXiv.org (Dec.\ 2003).

\bibitem{DuGP97} D. Du, J. Gu, P. Pardalos (Eds.),
{\it Satisfiability Problem: Theory and Applications.}
DIMACS Series in Discr.\ Math.\ and Theoret.\ Comput.\ Sci. 35,
American Math.\ Soc., Providence RI, 1997.

\bibitem{Duec93} G. Dueck, New optimization heuristics:
the great deluge algorithm and the record-to-record travel.
{\it J. Comput.\ Phys.\ 104} (1993), 86--92.

\bibitem{GaJo79} M. R. Garey and D. S. Johnson, {\it Computers and 
Intractability: A Guide to the Theory of NP-Completeness.} 
W. H. Freeman \& Co., San Francisco CA, 1979.

\bibitem{Gu92} J. Gu,
Efficient local search for very large-scale satisfiability problems.
{\it SIGART Bulletin 3:1} (1992), 8--12.

\bibitem{Hoos02} H. H. Hoos,
An adaptive noise mechanism for WalkSAT.
{\it Proc.\ 18th Natl.\ Conf.\ on Artificial Intelligence (AAAI-02)}, 
655--660. AAAI Press, San Jose Ca, 2002.

\bibitem{HoSt99} H. H. Hoos and T. St\"utzle,
Towards a characterisation of the behaviour of stochastic local
search algorithms for SAT.
{\it Artificial Intelligence 112} (1999), 213--232.

\bibitem{MaMW04} E. Maneva, E. Mossel and M. J. Wainwright,
A new look at survey propagation and its generalizations.
Technical report cs.CC/0409012, arXiv.org (April 2004).

\bibitem{Metr53} N. Metropolis, A. Rosenbluth, M. Rosenbluth,
A. Teller and E. Teller, 
Equations of state calculations by fast computing machines, 
{\it J. Chem. Phys. 21} (1953), 1087--1092.

\bibitem{MeRZ03} M. M\'ezard, F. Ricci-Tersenghi, and R. Zecchina, 
Alternative solutions to diluted p-spin models and XORSAT problems. 
{\it J. Stat. Phys. 111} (2003), 505--533.

\bibitem{MeZe02} M. M\'ezard and R. Zecchina,
Random K-satisfiability  problem: From an analytic solution
to an efficient algorithm.
{\it Phys. Rev. E 66} (2002), 056126.

\bibitem{MiSL92} D. Mitchell, B. Selman and H. Levesque,
Hard and easy distributions of SAT problems.
{\it Proc.\ 10th Natl.\ Conf.\ on Artificial Intelligence (AAAI-92)}, 
459--465. AAAI Press, San Jose CA, 1992.

\bibitem{MoPR04} A. Montanari, G. Parisi and F. Ricci-Tersenghi,
Instability of one-step replica-symmetry-broken phase in 
satisfiability problems.
{\it J. Phys. A 37} (2004), 2073--2091.

\bibitem{Papa91} C.H. Papadimitriou,
On selecting a satisfying truth assignment.
{\it Proc.\ 32nd IEEE Symposium on the Foundations of Computer
Science (FOCS-91)}, 163--169.
IEEE Computer Society, New York NY, 1991.

\bibitem{Pari02} G. Parisi,
On local equilibrium equations for clustering states.
Technical report cs.CC/0212047, arXiv.org (Feb 2002).

\bibitem{Park01} A. J. Parkes,
Distributed local search, phase transitions, and polylog time.
{\it Proc.\ Workshop on Stochastic Search Algorithms,
17th International Joint Conference on Artificial Intelligence
(IJCAI-01)}. 2001.

\bibitem{Park02} A. J. Parkes,
Scaling properties of pure random walk on random 3-SAT.
{\it Proc.\ 8th Intl.\ Conf.\ on Principles and Practice
of Constraint Programming (CP 2002)}, 708--713.
Lecture Notes in Computer Science 2470.
Springer-Verlag, Berlin 2002.

\bibitem{SeOr03} S. Seitz and P. Orponen,
An efficient local search method for random 3-satisfiability.
{\it Proc.\ IEEE LICS'03 Workshop on Typical Case Complexity
and Phase Transitions}. Electronic Notes in Discrete
Mathematics 16, Elsevier, Amsterdam2003.

\bibitem{SeKC96} B. Selman, H. Kautz, B. Cohen,
Local search strategies for satisfiability testing.
In: D. S. Johnson and M. A. Trick (Eds.), 
{\it Cliques, Coloring, and Satisfiability}, 521--532.
DIMACS Series in Discr.\ Math.\ and Theoret.\ Comput.\ Sci. 26,
American Math.\ Soc., Providence RI, 1996.

\bibitem{SeLM92} B. Selman, H. J. Levesque and D. G. Mitchell,
A new method for solving hard satisfiability problems.
{\it Proc.\ 10th Natl.\ Conf.\ on Artificial Intelligence (AAAI-92)}, 
440--446. AAAI Press, San Jose CA, 1992.

\bibitem{SeMo03} G. Semerjian and R. Monasson,
Relaxation and metastability in a local search procedure
for the random satisfiability problem.
{\it Phys.\ Rev.\ E 67} (2003), 066103.

\bibitem{SeMo04} G. Semerjian and R. Monasson,
A study of Pure Random Walk on random satisfiability
problems with ``physical'' methods.
{\it Proc.\ 6th Intl.\ Conf.\ on Theory and Applications
of Satisfiability Testing (SAT 2003)}, 120--134.
Lecture Notes in Computer Science 2919.
Springer-Verlag, Berlin 2004.

\bibitem{WeSe02} W. Wei and B. Selman,
Accelerating random walks.
{\it Proc.\ 8th Intl.\ Conf.\ on Principles and Practice
of Constraint Programming (CP 2002)}, 216--232.
Lecture Notes in Computer Science 2470.
Springer-Verlag, Berlin 2002.

\end{thebibliography}
\end{document}